\documentclass[11pt]{article}

\usepackage[utf8]{inputenc}
\usepackage[T1]{fontenc}
\usepackage{lmodern}
\usepackage{amsmath, amssymb, amsthm}
\usepackage{graphicx}
\usepackage[table]{xcolor}
\usepackage{booktabs}
\usepackage{siunitx}
\usepackage{float}

\usepackage{subcaption}
\usepackage{float}
\usepackage{hyperref}
\usepackage{geometry}
\usepackage{cite}
\usepackage{authblk}
\usepackage{titlesec}
\usepackage{colortbl}
\usepackage{tikz}
\usetikzlibrary{fit}
\definecolor{mydiag}{HTML}{21918C}

\titleformat{\section}{\large\bfseries}{\thesection.}{0.5em}{}
\titleformat{\subsection}{\normalsize\bfseries}{\thesubsection.}{0.5em}{}

\title{Geometric Features of Higher-Order Networks via the Spectral Triplet}
\author[1,2]{Sara Najem}
\author[1]{Dima Mrad}
\author[2]{Mohammad Elsayed}
\affil[1]{\small Center for Advanced Mathematical Sciences, American University of Beirut, Beirut, Lebanon}
\affil[2]{\small Department of Physics, American University of Beirut, Beirut, Lebanon}
\date{}

\begin{document}

\maketitle
Our work is concerned with simplicial complexes that describe higher-order interactions in real complex systems. This description allows to go beyond the pairwise node-to-node representation that simple networks provide and to capture a hierarchy of interactions of different orders.
The prime contribution of this work is the introduction of geometric measures for these simplicial complexes. We do so by noting the non-commutativity of the algebra associated with their matrix representations and consequently we bring to bear the spectral triplet formalism of Connes on these structures and then notions of associated dimensions, curvature, and distance can be computed to serve as characterizing features in addition to known topological metrics. We demonstrate the formalism on a real data set of musical composition and reveal latent geometric structures of these composition.

\begin{abstract}
\end{abstract}
\section{Introduction}
Networks and higher-order networks, including hypergraphs and simplicial complexes, have become essential in modeling complex systems. Higher-order networks, in particular, have gained attention over the past few years in the study of systems that exhibit inherently multiscale interactions \cite{battiston2021physics,bianconi2021higher,benson2016higher,scholtes2016higher,mayfield2017higher,lambiotte2019networks,battiston2022higher}. The latter, which are in this framework higher-order edges, can be represented as complexes or faces of a simplicial complex whose characteristics are fully understood in algebraic topology.  These characteristics are nothing but higher-order generalizations of the topological metrics of simple networks like degree, centrality, and others. The fact that these structures can be studied in an algebraic topological framework begs the question of the link to discrete differential geometry and subsequently to the possibility of revealing geometrical features that are hidden therein. 

The correspondence between geometry and algebra is established through the concept of the spectral triplet, introduced by Connes in the context of non-commutative geometry. It offers a powerful framework to generalize classical geometric concepts to settings where the underlying space is non-commutative \cite{connes1995local,connes2010noncommutative,connes2013spectral,khalkhali2008invitation}. The spectral triplet encodes metric, topological, and differential information through spectral data. This framework naturally accommodates discrete structures that are not necessarily manifolds, which makes it convenient for extending geometric notions, such as curvature, to higher-order networks. 
Of particular interest to us here are dimension, distance and curvature. In recent years, the latter has gained attention in the complex sciences community, with multiple definitions to compute it over complex networks and their higher-order counterparts. These definitions vary from purely combinatorial ones, such as the Forman-Ricci curvature \cite{forman2003bochner,leal2021forman,sreejith2016forman,weber2017characterizing}, to optimal transport, such as the Ollivier-Ricci curvature \cite{ollivier2009ricci,bauer2011ollivier,lin2011ricci} as well as other methods rooted in geometric embedding of networks \cite{li2023hyperbolic}. As for a definition of distance over higher order networks, the route is less traveled and there is no consensus of how to define it \cite{vasilyeva2023distances,nortier2025higher,chakraborty2025shortest}. As for dimension, its definition is somewhat agreed upon \cite{torres2020simplicial,burioni1996universal,hwang2010spectral,durhuus2007spectral}, which we adopt in this work and put to use in the definition of curvature.  In this paper, we use the spectral action principle and the heat kernel expansion and borrow the spectral definition of curvature over these networks from the continuous counterpart.  In addition, we use Connes' definition of distance, which somewhat does away with notion of "points" in classic geometrical spaces, and find the spacing between the musical notes in the non-commutative space represented by their higher order networks. Also, inspired by Weyl's law governing the spectrum of the Laplace-Beltrami we operator here we use that of the Hodge Laplacian to infer the dimension of the spaces in which these compositions are embedded.

\section{Structure of Higher-Order Networks}

\subsection{Higher-order networks}
A Higher-Order Network $\mathcal{N}$ is composed of a set of vertices $\mathcal{V}$ and hyperedges $\mathcal{H}$ and is represented as $\mathcal{N} = (\mathcal{V}, \, \mathcal{H})$. One class of these higher-order networks are simplicial complexes with  simplices as their building block. A simplex of dimension $k$ is a set of $k+1$ interacting nodes,$\alpha = \left[v_0, v_1, v_2, ..., v_k\right]$ with a defined orientation. These higher-order node interactions consider simultaneous interactions between more than two interacting entities of the system \cite{bianconi2021higher,millan2022geometry,torres2020simplicial}.

\subsection{Boundary operator and Incidence matrices}
In the context of higher-order networks, the Boundary operator is used to generalize incidence matrices employed in graph theory. The entries of the incidence matrices are binary values that indicate whether lower-dimensional simplices are included in higher-dimensional ones. 
These incidence matrices are computed as follows, $\left[B_{[n]}\right]_{\alpha ', \alpha} =  (-1)^p$, for every pair of $n$- dimensional simplex $\alpha$ and $(n-1)$- dimensional simplex $\alpha '$ such that: $\alpha = [v_0, v_1, ..., v_n]; \ \alpha' = [v_0, v_1, ..., v_{p-1}, v_{p+1}, ..., v_n]$ \cite{bianconi2021higher,millan2022geometry,torres2020simplicial,millan2025topology}.

\subsection{Higher-order Hodge Laplacian and Super-Laplacian}
The traditional graph Laplacian defined as $L = D - A$ can be alternatively written as
$L = BB^T$, where $B$ is the Boundary (Incidence) matrix. For a $k$-simplicial complex, the $n$-order Hodge Laplacian is represented by the matrix \cite{bianconi2021higher,torres2020simplicial}:

\begin{equation}\label{eq:Laplacian}
	L_{[n]} = B_{[n]}^T B_{[n]} + B_{[n+1]} B_{[n+1]}^T = L_{[n]}^{\text{down}} + L_{[n]}^{\text{up}}= \delta_{[n-1]} d_{[n]} + d_{[n+1]} \delta_{[n]}
\end{equation}

The super Laplacian $L_s$ is defined as a block diagonal matrix composed of the Hodge Laplacians of all orders:

\begin{equation}\label{lsup}
L_s = \begin{pmatrix}
L_{[0]} & 0 & 0 & 0 \\
0 & L_{[1]} & 0 & 0 \\
0 & 0 & L_{[2]} & 0 \\
0 & 0 & 0 & L_{[3]} \\
\end{pmatrix}
\end{equation}
\\
where $L_k = B_k B_k^\top + B_{k+1}^\top B_{k+1}$ is the $k$-th order Hodge Laplacian for $k=0,1,2,3$.\\


\subsection{Dirac Operator}

Let $ B_m \in \mathbb{R}^{n_{m-1} \times n_m} $ denote the incidence matrix between $(m-1)$-simplices and $m$-simplices in a $k$-dimensional simplicial complex.

The Dirac operator $ D $ \cite{bianconi2021topological,calmon2022higher,calmon2023dirac} defined as the square root of the super-Laplacian, acts on the space of all $k$-simplices for $k = 0, 1, \dots, d$, and takes the following block matrix form:

$$
D = L^{\frac{1}{2}} = 
\begin{pmatrix}
0 & B_1 & 0 & \cdots & 0 & 0 \\
B_{[1]}^\top & 0 & B_{[2]} & \cdots & 0 & 0 \\
0 & B_{[2]}^\top & 0 & \cdots & 0 & 0 \\
\vdots & \vdots & \vdots & \ddots & \vdots & \vdots \\
0 & 0 & 0 & \cdots & 0 & B_{[k]} \\
0 & 0 & 0 & \cdots & B_{[k]}^\top & 0
\end{pmatrix}
$$


\section{Geometry}\label{sec:spectrip}
To motivate the relation between geometry and spectra in our context, we go back to Kac's question on whether we can "hear the shape of the drum". For a few aspects of the geometry the answer is the affirmative.  For example, Weyl's law \cite{connes2010noncommutative,buser2010geometry} reveals the dimension of the manifold through the dependence of $N$, the number of eigenvalues of the Laplace-Beltrami operator on $\lambda$ in the asymptotic limit. It is given by $N(\lambda \leq \lambda_i) \propto \lambda ^{n/2}$, where $n$ is the dimension. What about the discrete setting? Is there any equivalent notion for dimension that can be inferred from spectra?

\subsection{Dimension}
In the context of higher-order networks, particularly simplicial complexes, the spectral dimension is not adequately described by a single scalar quantity given the complex geometry of these structures. Instead, for a simplicial complex of dimension $k$, a vector of spectral dimensions \cite{bianconi2021higher} characterizes the geometry : $$d_s = [d_s^{[0]}, d_s^{[1]}, d_s^{[2]}, \cdots, d_s^{[k-1]}]$$
where $d_s^{[0]}$ corresponds to the spectral dimension associated with the Laplacian $L_0$ of a regular pairwise graph, and every other component is associated with the $n-$order up-Laplacian $L^{[n]}_{\text{up}}=B_{[n+1]} B_{[n+1]}^T$ governing diffusion processes across $m-$simplices. 
For the graph Laplacian, the first component of the spectral dimension vector, the scaling of the cumulative density of the non zero eigenvalues \cite{torres2020simplicial} is given by :
$$\rho(\lambda) = C_{[0]} \lambda^{d^{[0]}_S / 2 }, \quad \text{for } n = 0, \\$$
where $C_{[0]}$ is a constant and $d^{[0]}_S$ is the spectral dimension of a graph or the $1-$skeleton of the simplicial complex. For higher-order up-Laplacians, the proportionality remains valid (check section \ref{lapop} in the Appendix). Specifically, for \(0 < n < k\), the spectral density of the \(n\)-order Laplacian is composed of two contributions,
$$
\rho(\lambda) = \frac{\mathcal{N}_{[n-1]}}{N_{[n]}} C_{[n-1]} \lambda^{d^{[n-1]}_S / 2 } + \frac{\mathcal{N}_{[n]}}{N_{[n]}} C_{[n]} \lambda^{d^{[n]}_S / 2 },
$$
where $\mathcal{N}$ is the number of non-zero eigenvalues of the Laplacian of a given order and $N$ is the number of simplices of the same order.

At the upper boundary, when \(n = k\), only the \((k-1)\)-simplices contribute, and the density simplifies to:
$$
\rho(\lambda) = C_{[k-1]} \lambda^{d^{[k-1]}_S / 2 }.
$$
\subsection{Spectral Triplet}
What other information can we extract from the spectra? The answer is now with Connes and his spectral triplet $(\mathcal{A},\mathcal{H},D)$ \cite{connes1995local,connes2010noncommutative,connes2013spectral,khalkhali2008invitation,khalkhali2013basic} in which a non-commutative geometry is associated with an non-commutative algebra $\mathcal{A}$, and a Hilbert space $\mathcal{H}$, in analogy with classical geometry's correspondence to algebra on smooth manifolds.
Why is this non-commutative framework relevant to our setting? Clearly, the algebra $\mathcal{A} = M_m(\mathbb{C})$ is the set of all $m\times m$ matrices endowed with addition, multiplication, conjugation, and scalar multiplication, which is obviously non-commutative (for example, $A$ and $B \in \mathcal{A}$ then generally $AB \neq BA$), which is the set of higher-order networks representing our system of interest. 
As the elementary building blocks of our networks' Hilbert spaces, we chose $\{v_i\}$, $\{e_{ik}\}$, $\{f_{ijk}\}$, $\{t_{ijkl}\}$ corresponding to the vertex, edge, triangle, tetrahedron sets
as basis elements of a certain hierarchy of Hilbert spaces over $\mathbb{C}$  with a scalar product induced by $(v_i |v_k ) = \delta_{ik}$ and $(e_{ik} |e_{lm} ) =\delta_{il} \delta_{km}$, which carries over to higher-order edges, like triangles and tetrahedra.

$$
\mathcal{H} = \begin{pmatrix}
\mathcal{H}_0 \\
\mathcal{H}_1\\
\mathcal{H}_2\\
\mathcal{H}_3\\
..
\end{pmatrix}
 = \mathcal{H}_0 \oplus \mathcal{H}_1 \oplus\mathcal{H}_2  \oplus \mathcal{H}_3...
$$

Thus $M_m(\mathbb{C})$ act on the Hilbert $\mathcal{H}$ space via matrix multiplication. As for the Dirac operator the previous section illustrated how to construct it in our discrete setting. Now we have all the ingredients to extract two very important additional geometrical measures: distance and curvature.

\subsubsection{Distance \`a la Connes}
In non-commutative geometry Connes introduced a notion of distance through functionals $\phi$ and $\psi$ which act on the algebra $\mathcal{A}$ and generalize the notion of points in classical spaces.  \cite{connes1995local,connes2010noncommutative,connes2013spectral}.   These functionals are related to the geometric information encoded in the spectrum of $D$ through the following equation that defines the distance $d_C$:
$$
d_C(\phi, \psi) = \sup_{a \in \mathcal{A}} \big\{ |\phi(a) - \psi(a)| \,\big|\, \|[D, a]\| \leq 1 \big\}
$$
Surprisingly, the geodesic distance in commutative spaces over Riemanian manifolds can be recovered from this definition. 

In the discrete setting on the other hand, this notion of distance has been explored in lattices and graphs/networks \cite{requardt2000graph,dimakis1998connes,besnard2021estimating,requardt1997new}. In a network setting, it was shown in \cite{requardt2002dirac} that the distance between two nodes $i$ and $j$ can be written as: 
$$
\mathrm{dist}_{C,\mathcal{G}}(i, j) := \sup \{ |f_j - f_i| \;:\; \|[D, f]\| \leq 1 \},
$$
where $f$ is  a \text{$0$-form} (or equivalently a scalar) defined on the nodes, whose indices are the subscript of $f$. The subscript $\mathcal{G}$ denotes that the distance computation is carried over simple networks. Below is our generalized definition for a simplicial complex $\mathcal{N}$:

\begin{equation}\label{eqsimplex}
\mathrm{dist}_{C,\mathcal{N}}(i, j) := \sup \{ |f_j - f_i| \;:\; \|[D, f]\| \leq 1 \}, 
\end{equation}
 which differs from that of $\mathcal{G}$ by  additional higher-order constraints which will be enumerated shortly. 

As given in \cite{requardt2002dirac}, we start by noting that: $$
\|[D, f]\| = \|[B, f]\|
.$$Subsequently, the commutator for the lowest order is now given: 
\begin{equation}\label{0form}
[B_0, f]\alpha(i,j) = \frac{(f_i - f_j)}{2}(\alpha(i) + \alpha(j)) 
\end{equation}
We note that equation \ref{0form} allows us to the write the constraint $\|[B_0, f]\| \leq 1$ as: 
\begin{equation}\label{0formcon}
\|[D, f]\|= \|\frac{(f_i - f_j)}{2}\| \leq 1.
\end{equation}
The details of this calculation are given in the Appendix along with the higher order generalizations in section \ref{const}. 

Now we list the higher-order constraints that are not imposed in the formulation of distance over $\mathcal{G}$. Similarly to what we did for an edge, for an oriented triangle, $[i,j,k]$ the equation for the commutator is given by: 
\begin{equation}\label{1form}
{
([B_1,f]\omega)([i,j,k]) = 
(f_i - f_j)\omega([j,k]) +
(f_j - f_k)\omega([k,i]) +
(f_k - f_i)\omega([i,j]).
}
\end{equation}
Finally, for an oriented tetrahedron $[i,j,k,l]$, the commutator is given by: 
\begin{equation}\label{2form}
{
\begin{aligned}
([B_2,f]\beta)([i,j,k,l]) &= 
(f_i - f_j)\,\beta([j,k,l]) +
(f_j - f_k)\,\beta([k,l,i]) \\
&\quad +
(f_k - f_l)\,\beta([l,i,j]) +
(f_l - f_i)\,\beta([i,j,k]).
\end{aligned}
}
\end{equation}

If we were to write equations \ref{1form} and \ref{2form} and sum over the elements of the simplex we observe that it telescopes to zero (for example, the triangle's constraint is given by: $f_i - f_j + f_j - f_k + f_k - f_i = 0 $) leading to  $\|[B_1, f]\| = 0$ and $\|[B_2, f]\| = 0$. Therefore, we enforce the constraint on all pairwise interactions in the simplex $\|\frac{f_i - f_j}{2}\| \leq 1$,  $\|\frac{f_j - f_k}{2}\| \leq 1$,  $\|\frac{f_k - f_i}{2}\| \leq 1$. This carries over to the tetraherda and higher-order simplices. 

Therefore in order to compute the distance between two nodes in the non-commutative space associated with the simplicial complex the above optimization problem given in equation \ref{eqsimplex} has to be solved with the constraints given in Equation \ref{0formcon}, and for each pair appearing in Equations \ref{1form}, and \ref{2form}.

\subsubsection{Spectral Action: the Heat Kernel and Expansion and Scalar Curvature}
Now we turn to the last piece of the puzzle and remind ourselves that for a general integral operator the kernel is given by: 
$$
(A_K f)(x) = \int K(x, y)\, f(y)\, dy
$$
whose trace is given by: 
$$
\operatorname{Tr}(A_K) = \int K(x, x)\, dx = \sum \sigma_i,
$$
where the $\sigma_i$ are the eigenvalues of the operator. 
Chamseddine and Connes's spectral action principle \cite{chamseddine1997spectral} states that the trace of any function $F$ of the Dirac operator can be used to infer information about curvature. The action is given by: 
\begin{equation*} 
    \mathrm{Tr}\left(F\left(\frac{D}{\Lambda}\right)\right) = \sum_i F\left(\frac{\lambda_i}{\Lambda}\right),
\end{equation*}
where $ \{ \lambda_i \} $ are the eigenvalues of $ D $, and
$ds = D^{-1} = |1 /\Lambda_1|  = 1/ \Lambda$, where $\Lambda_1$ is the smallest non-zero eigenvalue of $D$. 
In the commutative case over ordinary manifolds this spectral action principle was shown to recover the Einstein–Hilbert action for gravity, Yang–Mills actions for gauge fields, and scalar field terms (like the Higgs field). In certain non-commutative geometries, applying the spectral action reproduces the Standard Model of particle physics coupled to gravity. 
Thus a suitable function $F$ of the Dirac operator can be related to its kernel and what kernel is better understood than the heat kernel?
Therefore, they choose to solve the heat equation on a Riemannian manifold \( (M, g) \): 
\begin{equation}\label{heatequ}
\frac{\partial c}{\partial t} = -\Delta c, 
\end{equation}
with $c_0 = \delta( x- y)$. 
Its solution is given by: 
\begin{equation} \label{heatc0}
c(x, t) = \int K(x,y,t) c_0 dy = \int K(x,y,t) \delta(x-y) dy = K(x,x,t)\end{equation}
Explicitly, the heat kernel has the following short-time asymptotic Minakshisundaram–Pleijel  expansion \cite{minakshisundaram1949some,chavel1984eigenvalues}:
\begin{equation}\label{heatkern}
K(x, y, t) \sim \frac{1}{(4\pi t)^{n/2}} \exp\left( -\frac{d(x,y)^2}{4t} \right) \sum_{k=0}^\infty u_k(x, y) t^k,
\end{equation}
and whose heat trace is given by:

$$
\operatorname{Tr}(e^{-t\Delta}) = \int K(x, x, t) \, dx = \frac{1}{(4\pi t)^{n/2}} \left( V(M) + \frac{t}{6}\int RdV+ \mathcal{O}(t^2)\right), $$
with $V(M)$ being the volume of the manifold and $R$ its scalar curvature.

Analogously, and in the context of higher-order networks, and in particular for simplicial complexes, the spectral action can be employed to obtain information about their curvatures. Here, we consider the combinatorial Dirac operator $ D $, and its square $ L_s = D^2 $, the super Laplacian replacing $\Delta$ in Equation \ref{heatequ}. Subsequently, the kernel is given by $K(t) = e^{-L_st/\Lambda^2}$,  with the indices $i$ and $j$ running over the elements of the simplicial complex. 
We proceed with the following exact analytic computation of the kernel. We start by diagonalizing $ L_s =  V^T \Sigma V $, where $ \Sigma $ is the diagonal matrix of the eigenvalues, and thus we can write:
\begin{equation}\label{kexact}
  K(t) =   e^{-tD^2/\Lambda^2 } = e^{-t L_s/\Lambda^2 } = V^T e^{-t \Sigma/\Lambda^2} V
\end{equation}
Further, and to simplify, we focus on the diagonals of the kernel $K(i,i,t)$, where $d(x,x) = 0$ of Equation \ref{heatkern}, which reduces the expansions of Equation \ref{heatkern} to $d-$independent terms. 
Due to the block structure of $L_s$ given in Equation \ref{lsup}, $K(i,i,t)$ is defined on nodes, edges, and triangles . Thus, it can be split into their corresponding components: $K_v(i,i,t)$, $K_e(i,i,t)$, $K_f(i,i,t)$  given by: 
\begin{equation}\label{kn}
K_v(i,i,t) \approx u_{0,v}(i,i)\left(\frac{1}{4 \pi t}\right)^{d_s^{[0]}/2} + u_{1,v}(i,i) \left(\frac{1}{4 \pi t}\right)^{d_s^{[0]}/2} t + \mathcal{O}(t^{2 -d_s^{[0]}/2}),
\end{equation}
Likewise: 
\[
K_e(i,i,t) \approx u_{0,e}(i,i)\left(\frac{1}{4 \pi t}\right)^{d_s^{[1]}/2} + u_{1,e}(i,i) \left(\frac{1}{4 \pi t}\right)^{d_s^{[1]}/2} t + \mathcal{O}(t^{2 -d_s^{[1]}/2}),
\]

\[
K_f(i,i,t) \approx u_{0,f}(i,i)\left(\frac{1}{4 \pi t}\right)^{d_s^{[2]}/2} + u_{1,f}(i,i) \left(\frac{1}{4 \pi t}\right)^{d_s^{[2]}/2} t + \mathcal{O}(t^{2 -d_s^{[2]}/2}),
\]
Now, how can one compute $u_1$ or equivalently curvature? The answer lies in Equation \ref{kn}. 
Clearly, the fit of these against $(4 \pi t)^{-d_s/2}$ and $t (4 \pi t)^{-d_s/2}$ and higher order powers (with $d_s = d_s^{[0]}, d_s^{[1]},d_s^{[2]}$) yields the coefficients $u_0$ and $u_1$ for every order of the simplicial complex, that is: curvature at every vertex $v$, edge $e$, and face $f$. We note that rescaling $L_s \rightarrow L_s/\Lambda^2$, or equivalently $t \rightarrow t/\Lambda^2$, rescales $u_k  \rightarrow u_k \Lambda^{d_s-2k}$, which we have to account for when recovering the coefficients.  
We note that there is another route that one can take to get the left-hand side of equation \ref{kn}, which is to recover the solution numerically and subsequently the kernel, using Crank-Nicolson algorithm for example.  Similar to Equation \ref{heatc0}, the solution of the heat equation on the simplicial complex is obtained using the super Laplacian $L_s$ and $c_0 = I$, where $I$ is the identity. It is given by the kernel $K(i,j,t)$ acting on the initial condition $I$:
\begin{equation}\label{heatsimp}
\begin{split}
    c(i,j,t) &= K(i,j,t)I(i,i,t)  \\
    c(i,i,t) &= K(i,i,t)
\end{split}
\end{equation}
Equation \ref{heatsimp} can be inverted to obtain the kernel, which is nothing but $c(i,i,t)$ given our suitably chosen initial condition. 

Either way, and in order to recover curvature, we need to ensure that the solution is obtained in the vicinity of $t^+\rightarrow 0^+$, where the expansion of equation \ref{heatkern} holds.
Further, we note that the trace of equation \ref{kexact} gives the following: 
\begin{equation}\label{trace}
\begin{aligned}
    \mathrm{Tr}(e^{-t L/\Lambda^2 }) &= \sum_i \left(1 - t \lambda/\Lambda^2 + \frac{t^2}{2} (\lambda/\Lambda^2)^2 - \cdots \right) \\
                           &= \mathrm{Tr}(I) - t\, \mathrm{Tr}(\Sigma /\Lambda^2) + \frac{t^2}{2} \mathrm{Tr}(\Sigma/\Lambda^2)^2 - \cdots  
\end{aligned}
\end{equation}
Conversely, when we trace out the expansion of the kernel of equation \ref{heatkern}, we get:
\begin{equation}\label{taylor}
\mathrm{Tr}(e^{-t L/\Lambda^2 })  =  \frac{\Lambda^{d_s}}{(4\pi t)^{n/2}} \left( \int u_0(x,x) dV + t \Lambda^{-2} \int u_1(x,x) dV + \cdots\right).\end{equation}
We recall that this expansion is only valid for $t$ close to zero, so to reconcile Equation \ref{taylor} with Equation \ref{trace} we take $\frac{1}{(4\pi t)^{n/2}} \rightarrow 1$, which now defines $t^+ \rightarrow  0^+$, the time we should evolve the solution to, using our numerical scheme, which now gives an approximation of $K(i,i,t)$ that can be fitted against time following equation \ref{kn}.
The same procedure can be applied to compute $K_e(i,i,t)$ and $K_f(i,i,t)$. 
We note that the use of edge curvature has been coupled with the Ricci flow \cite{lin2011ricci,sia2019ollivier}, to detect communities. Explicitly, negatively curved edges are associated with edges between communities, while positively curved edges lie within communities. The fit of $K_e(i,i,t)$ against time, obviously gives us $u_{1,e}$ for every edge and subsequently the task of grouping nodes based on the sign of curvature of their associated edges will be the topic of another work, which relies on the definition of social balance and frustration \cite{aref2020modeling}, where the goal is to optimize a function that penalizes the mismatched edges within communities. 
In what follows we focus on $K_v(i,i,t)$

This proposed computation of curvature is rooted in differential geometry and functional analysis, which our work brings to use in higher-order networks. It will be contrasted with other measures currently used in the literature, to which we give an overview in section \ref{seccur}.

\section{Bochner-Weitzenb{\"o}ck Decomposition and Forman-Ricci Curvature}\label{seccur}

\begin{figure}[!htp]
    \centering
    \includegraphics[width=0.7\linewidth]{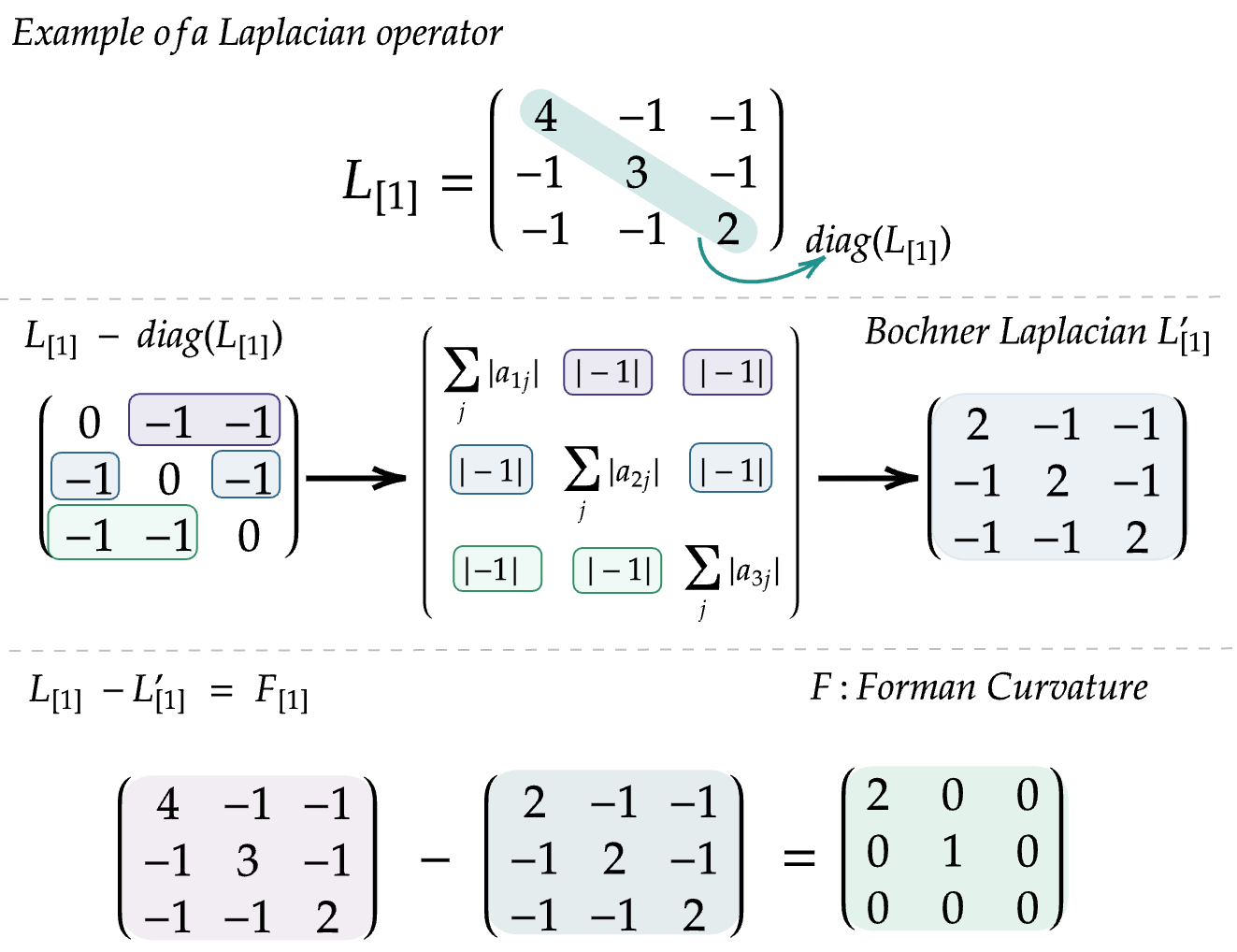}
    \caption{To compute the Forman edge-level curvature, we start with the Hodge Laplacian $L$ over edges, and construct the Bochner Laplacian $L'$ obtained by zeroing out the diagonal of $L$ and replacing it with the row-wise sum of the absolute value of the off-diagonal entries. The Forman curvature matrix is then obtained by subtracting $L'$ from $L$: $F = L - L'$. To compute vertex-level curvature, or scalar curvature, the curvature contribution of all edges incident to a vertex is summed.} 
    \label{fig:forman}
\end{figure}

An alternative definition of curvature over these structures is adapted from the Bochner-Weitzenb{\"o}ck Decomposition in a smooth setting \cite{forman2003bochner}, which decomposes the Hodge Laplacian $\Delta$ acting on differential forms as a sum of a \emph{rough} Laplacian and another term that depends on curvature:
$$
\Delta = \nabla^*\nabla + \mathcal{R}
$$
In this decomposition, $\Delta$ is the Hodge Laplacian, $\nabla$ is the covariant derivative, and $\mathcal{R}$ represents curvature.
A combinatorial analog \cite{sreejith2016forman,weber2017characterizing,leal2021forman} for this decomposition exists for simplicial complexes in discrete exterior calculus, based on Forman's construction. The discrete Hodge Laplacian of a $k-$dimensional simplicial complex then decomposes into two matrices: an off-diagonal matrix $L'_{[k]}$, representing the rough Laplacian, encoding diffusion across simplices, and a diagonal matrix $F_k$ encoding curvature. Formally, the decomposition is written as:
$$
L_{[k]} = L'_{[k]} + F_{[k]}
$$
An explicit example for computing the Forman curvature is provided and illustrated in Fig. \ref{fig:forman}.

Given a simplicial complex $\mathcal{N}$ with simplices of order n $\alpha^n \in \mathcal{N}$, the curvature at vertex $v \in \alpha^0$ is defined as an average of the Forman-Ricci curvature over all higher-order objects incident to the $v$:
\begin{equation}\label{formanvertex}
\mathrm{Ric}_F(v) = \frac{1}{\left| \{ \alpha \in \mathcal{N} : v \in \alpha \} \right|} \sum_{\substack{\alpha \in \mathcal{N} \\ v \in \alpha}} \mathrm{Ric}_F(\alpha)
\end{equation}

\newpage


\section{Case Study: Musical Data}
\begin{figure}[!htp]
    \centering
    \includegraphics[width=1\linewidth]{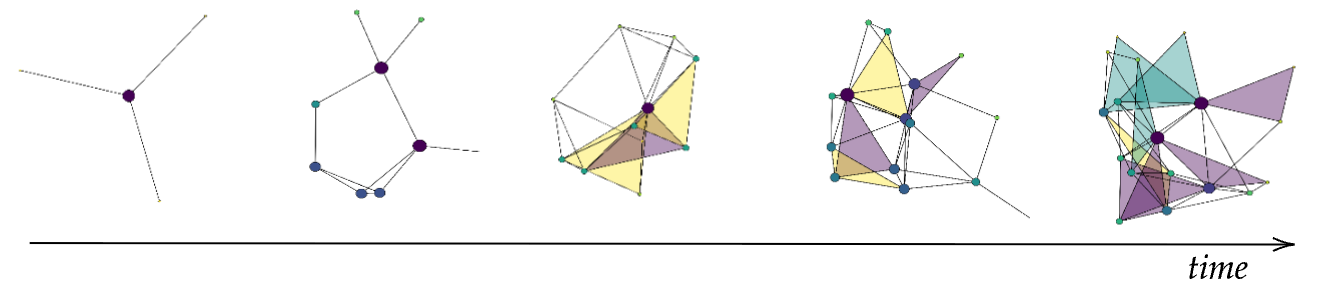}
    \caption{Visualization of the temporal evolution and cumulative construction of the simplicial complex}
    \label{fig:temporal-evolution}
\end{figure}

As a case study, we demonstrate the applicability of our approach by considering simplicial complexes constructed from musical data, particularly the sonatas and partitas for solo violin, composed by J. S. Bach. We group the movements of the sonatas and partitas into three categories based on their musical genre and character, the groups are slow movements, fugues, and dance movements. We construct the simplicial complexes corresponding to each of these movements. Notes are associated with nodes, or 0-simplices, chords of length two or two simultaneously played notes with a 1-simplex, a chord of length three or equivalently three simultaneously played notes with a triangle or a 2-simplex, all the way up to tetrahedron or a 3-simplex. These building blocks are now connected if they occur consecutively, that is if a $C$ is played followed by $[D,E,F]$, an edge is added between $C$, and the 2-simplex $[D,E,F]$. The complete details of the construction are given in \cite{mrad2025higher}. As a result, we obtain, for each movement, a simplicial complex of dimension $d=3$ with tetrahedra as the highest order simplex. We examine the temporal evolution of each simplicial complex by constructing it cumulatively, where at every musical measure $t$, new simplices are added to the complex, as the piece is played from start to end (illustrated in Fig. \ref{fig:temporal-evolution}).

\subsection{Dimension}
Here, we explore the asymptotic growth of the eigenvalues of the Hodge Laplacians of the musical movements to infer the associated dimension for each order. 
\begin{table}[H]
    \centering

    \caption{Values of dimensions $d_s^{[0]}$, $d_s^{[1]}$, and $d_s^{[2]}$ for different movements}
    \label{tab:spectraldims}
        \resizebox{0.4\textwidth}{!}{
    \begin{tabular}{@{}l S[table-format=2.2] S[table-format=2.2] S[table-format=2.2]@{}}
        \toprule
        \textbf{Movement} & \textbf{$d_s^{[0]}$} & \textbf{$d_s^{[1]}$} & \textbf{$d_s^{[2]}$} \\
        \midrule
        \textit{Partita 1 Allemanda}    & 2.706   & 2.940  & 5.310 \\
        \textit{Partita 1 Bourrée}      & 2.325  & 1.673   & 5.007 \\
        \textit{Partita 1 Sarabande}    & 2.026   & 2.286  & 9.194 \\
        \textit{Partita 2 Sarabande}    & 2.114   & 2.540  & 4.906  \\
        \textit{Partita 3 Loure}        & 2.400  & 2.057  & N/A  \\
        \textit{Partita 3 Minuet I, II} & 2.134  & 2.932  & N/A  \\

        \textit{Sonata 1 Adagio}        & 3.777  & 2.289  & 6.417 \\
        \textit{Sonata 1 Fugue}         & 1.933  & 1.666  & 2.209  \\        
        \textit{Sonata 1 Siciliano}     & 2.250  & 2.214  & 9.398 \\
        \textit{Sonata 2 Grave}         & 2.450   & 2.112  & 7.404  \\
        \textit{Sonata 2 Fugue}         & 3.792  & 1.648  & 6.195  \\
        \textit{Sonata 2 Andante}       & 2.209 & 2.025  & 4.792 \\
        \textit{Sonata 3 Adagio}        & 2.342  & 1.729  & 4.417  \\
        \textit{Sonata 3 Fugue}         & 2.180  & 1.713  & 4.081  \\
        \textit{Sonata 3 Largo}         & 2.161  & 4.065  & 3.645 \\
        \bottomrule
    \end{tabular}
    }
\end{table}
\subsection{Spectral curvature versus Forman-Ricci curvature}
\begin{figure}[H] 
        \centering
        \begin{minipage}[c]{0.49\textwidth} 
            \centering
        \includegraphics[width=\linewidth]{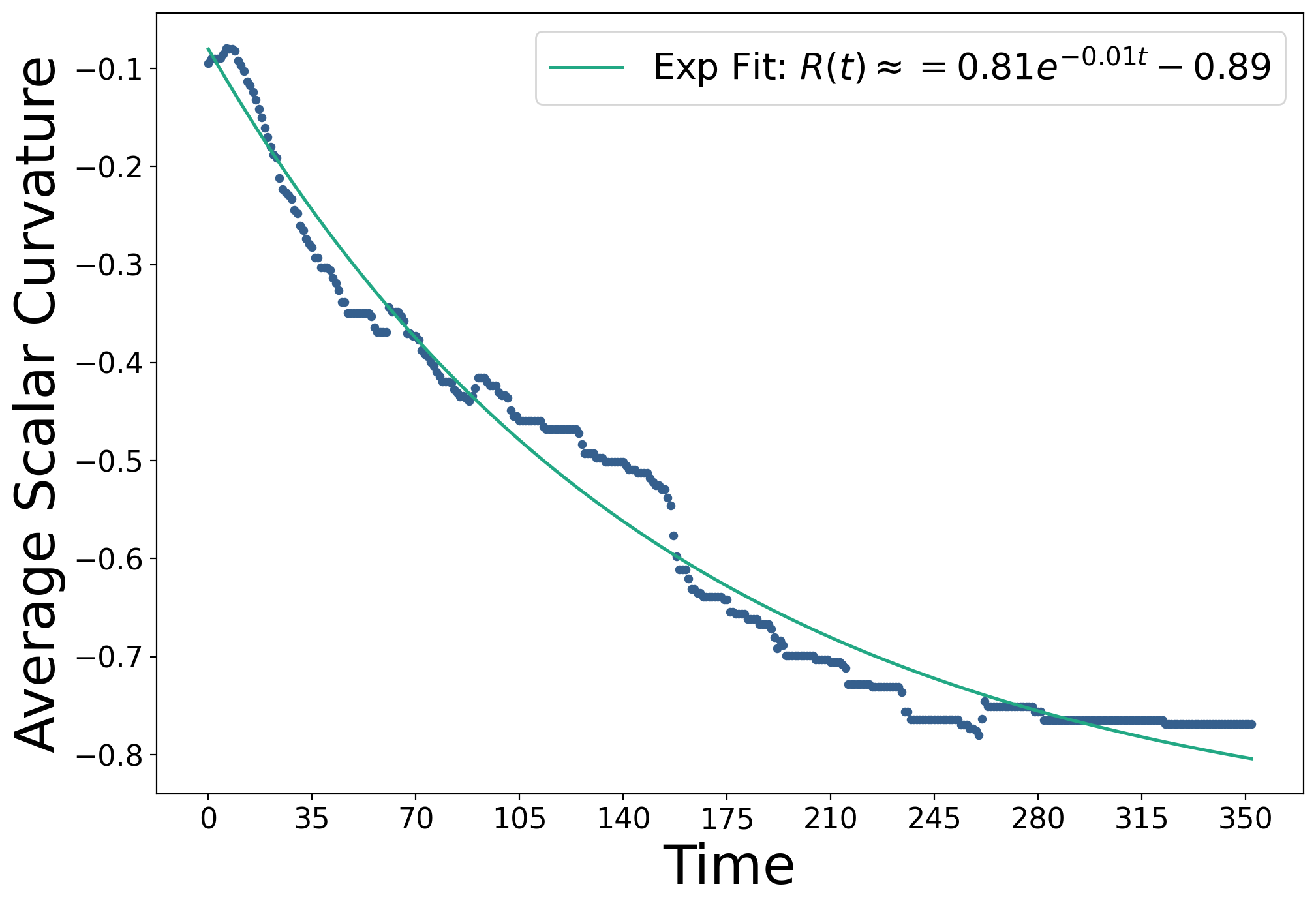}
            \caption{Evolution of the average scalar curvature using our proposed method for the Fugue from Sonata No. 3.}
            \label{fig:AvgScalarCurvFugue3}
        \end{minipage}
        \hfill
        \begin{minipage}[c]{0.49\textwidth} 
            \centering
        \includegraphics[width=\linewidth]{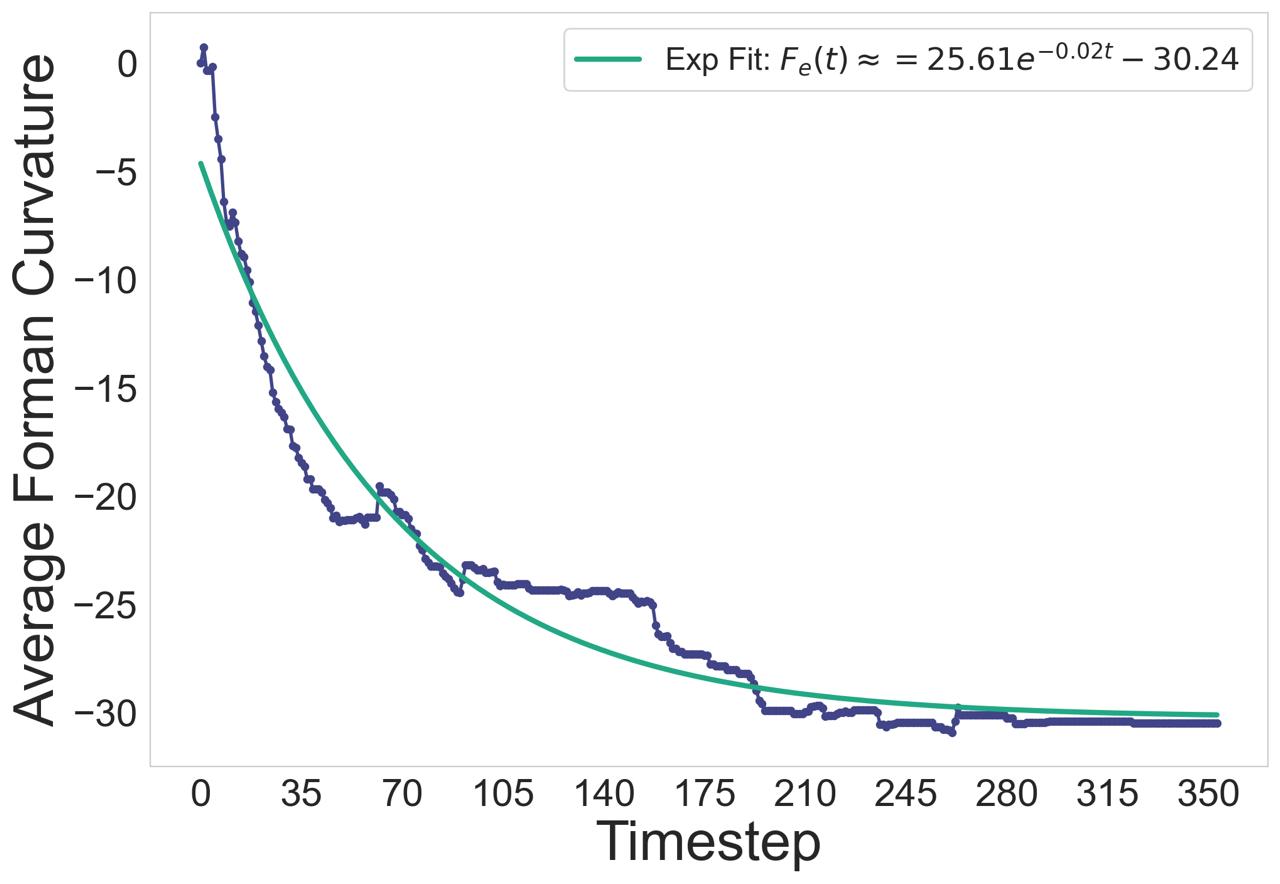}
            \caption{Evolution of the average Forman-Ricci node curvature for the Fugue from Sonata No. 3.}
            \label{fig:AvgFormanCurvFugue3}
        \end{minipage}
    \end{figure}

\begin{figure}[H]
\centering
    \begin{minipage}[t]{0.3\textwidth}
        \includegraphics[width=\linewidth]{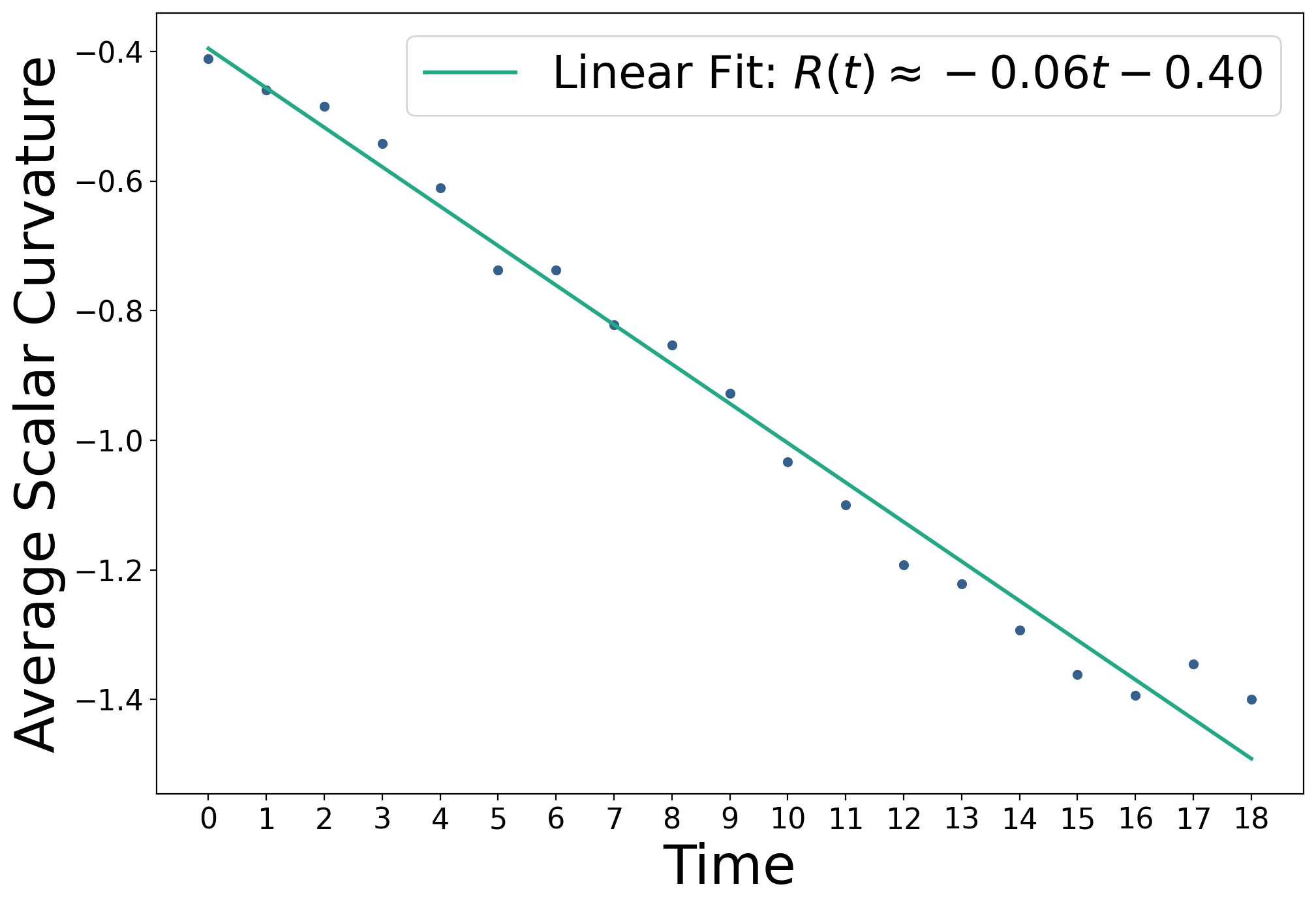}
        \caption{Average scalar curvature evolution using our proposed scalar curvature for slow movement Largo from Sonata 3 exhibits a linear trend.}
        \label{fit-sonata-3-largo}
    \end{minipage}
   \hspace{0.04 \textwidth}
    \begin{minipage}[t]{0.3\textwidth}
        \includegraphics[width=\linewidth]{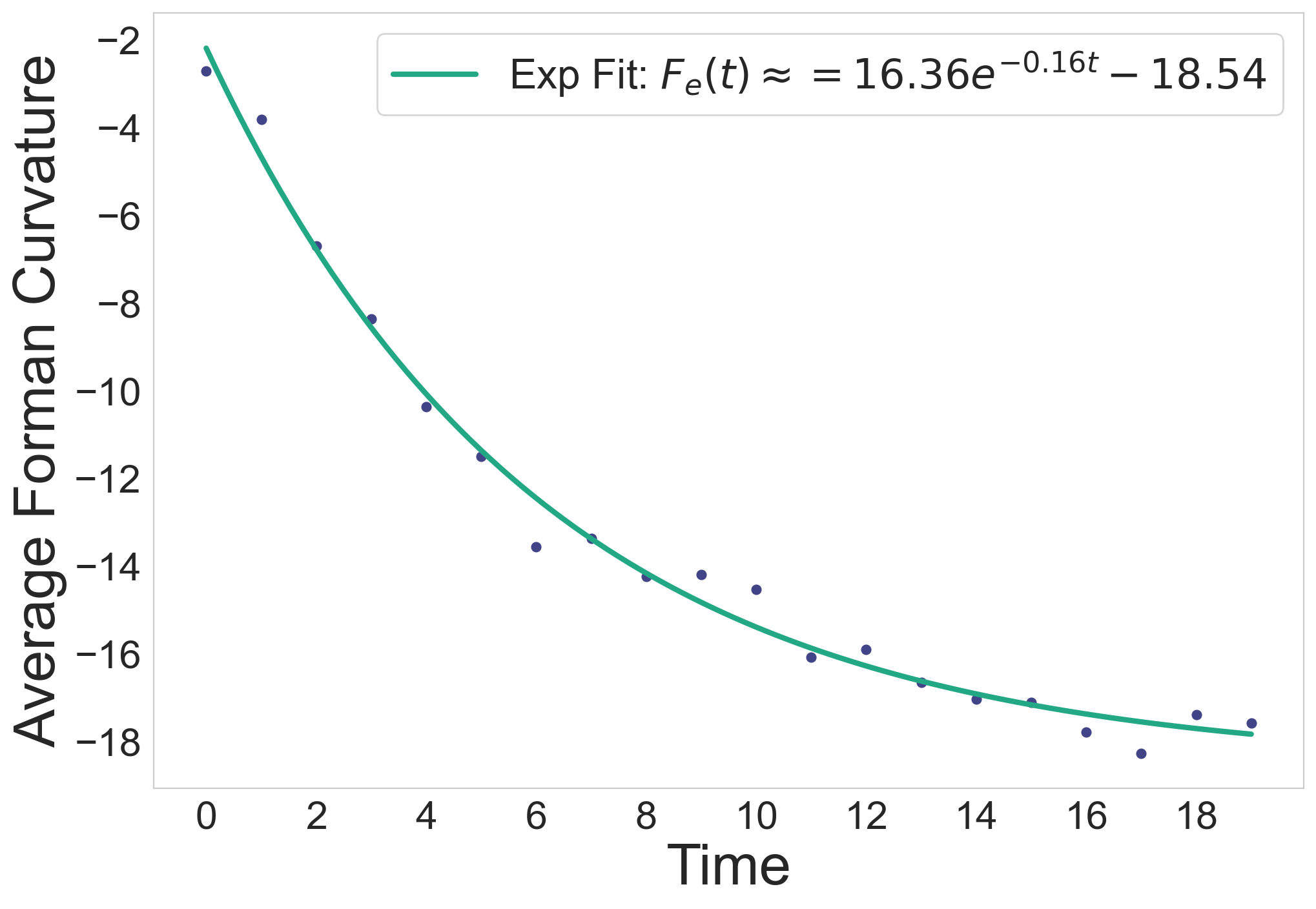}
        \caption{Average Forman-Ricci node curvature evolution for slow movement Largo from Sonata 3.}
        \label{forman-sonata-3-largo}
    \end{minipage}
    \end{figure}

We start by finding $\Lambda$ associated with $D$ or equivalently $L_s$ of the whole piece. The values of $\Lambda$ for each movement are tabulated in Table \ref{tab:lambdas} in the Supplementary Results section in the Appendix. This allows us to compute $K(i,i,t)$ using equation \ref{kexact} which we fit against the terms of equation \ref{kn}. 
truncating the expansion at $t^{4 - d_s^{[0]}/2}$ term, and subsequently recover $u_1$ from the fit at $t \rightarrow 0$ (explicitly up to $t = 1/(4 \pi )$).  This allows us to compute the evolution of the node-level curvatures of the movements using the spectral action principle and the heat-kernel expansion of $K(x,x,t)$. We then compare them with the Forman-Ricci  curvature. We have mentioned in Section \ref{seccur} that the Forman-Ricci curvature is an edge-based curvature measure. To compute it over vertices, it was considered reasonable to take the average of the hyperedges incident to the vertex, as shown in Eq. \ref{formanvertex}. We present below the temporal evolution of the individual note curvatures and compare them as computed by our proposed method against those obtained from the Forman-Ricci node curvature. In general, both approaches reflect broadly similar qualitative behavior in curvature dynamics and in the averaged curvature over nodes (Figs. \ref{fig:AvgFormanCurvFugue3} and \ref{fig:AvgScalarCurvFugue3}), however, we observe notable differences in how node-level behavior. The divergence in behavior across nodes is greater in our proposed method, which means that locally, the structural evolution is more diverse. In contrast, the behavior in the Forman-Ricci node-level curvature exhibits more similar trends across nodes.  By definition, the Forman curvature depends on local connectivity, 
making it sensitive to the degree of the node and its local fluctuations. The Forman curvature essentially captures a smoothed representation of degree evolving with time. Our proposed method however, computes curvature from a purely spectral geometric perspective. A sample of the results is shown in the Appendix in Figures \ref{lap-partita-2-sarabande},\ref{forman-partita-2-sarabande}, \ref{fit-sonata-3-largo}, and \ref{forman-sonata-3-largo}.  

    \begin{figure}[H] 
        \centering
        \begin{minipage}[c]{0.427\textwidth} 
            \centering
        \includegraphics[width=\linewidth]{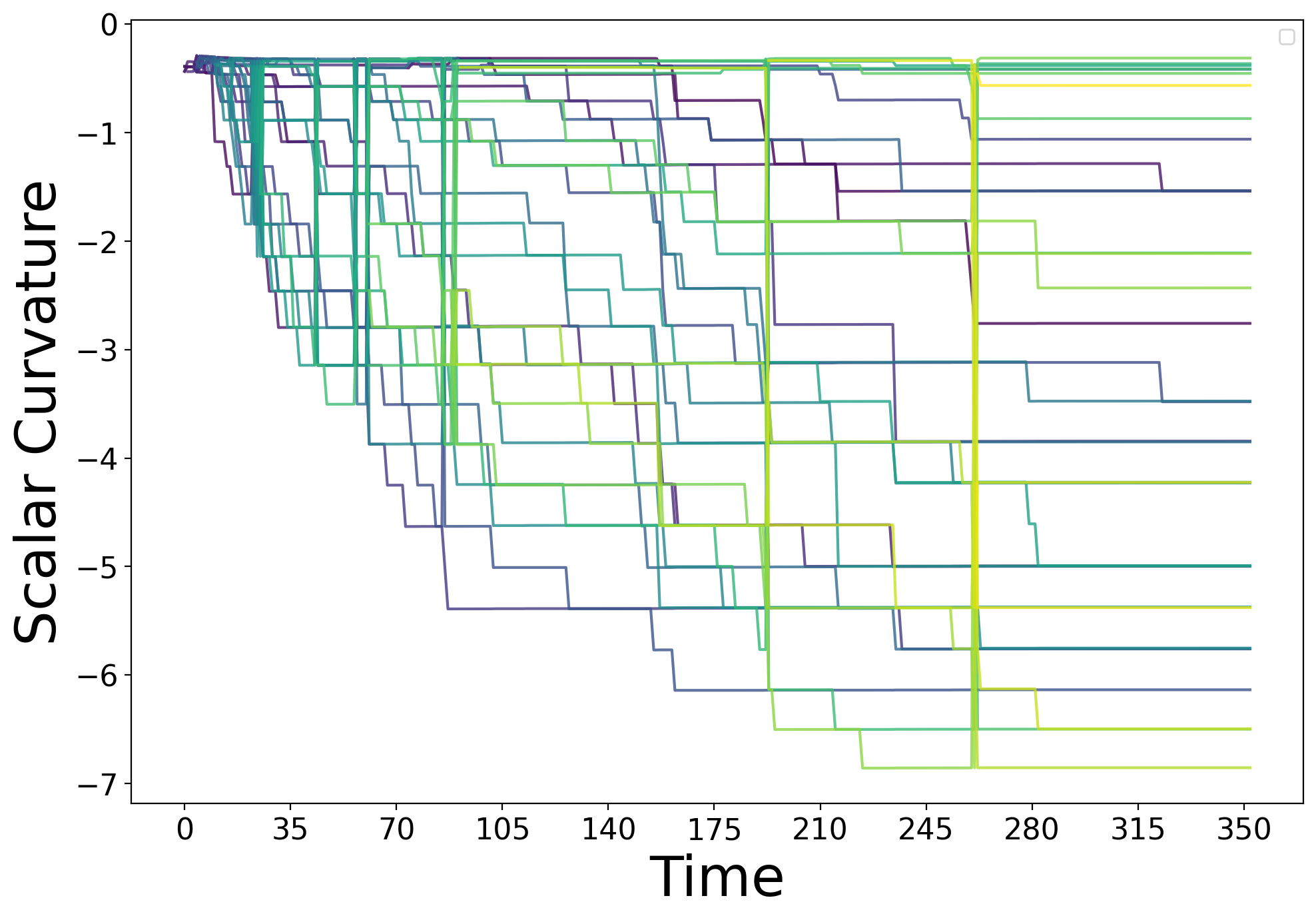}
            \caption{Evolution of the heat kernel expansion scalar curvatures over vertices for Fugue from Sonata No. 3}
            \label{fig:laplacianNodeCurv}
        \end{minipage}
        \hfill
        \begin{minipage}[c]{0.55\textwidth} 
            \centering
        \includegraphics[width=\linewidth]{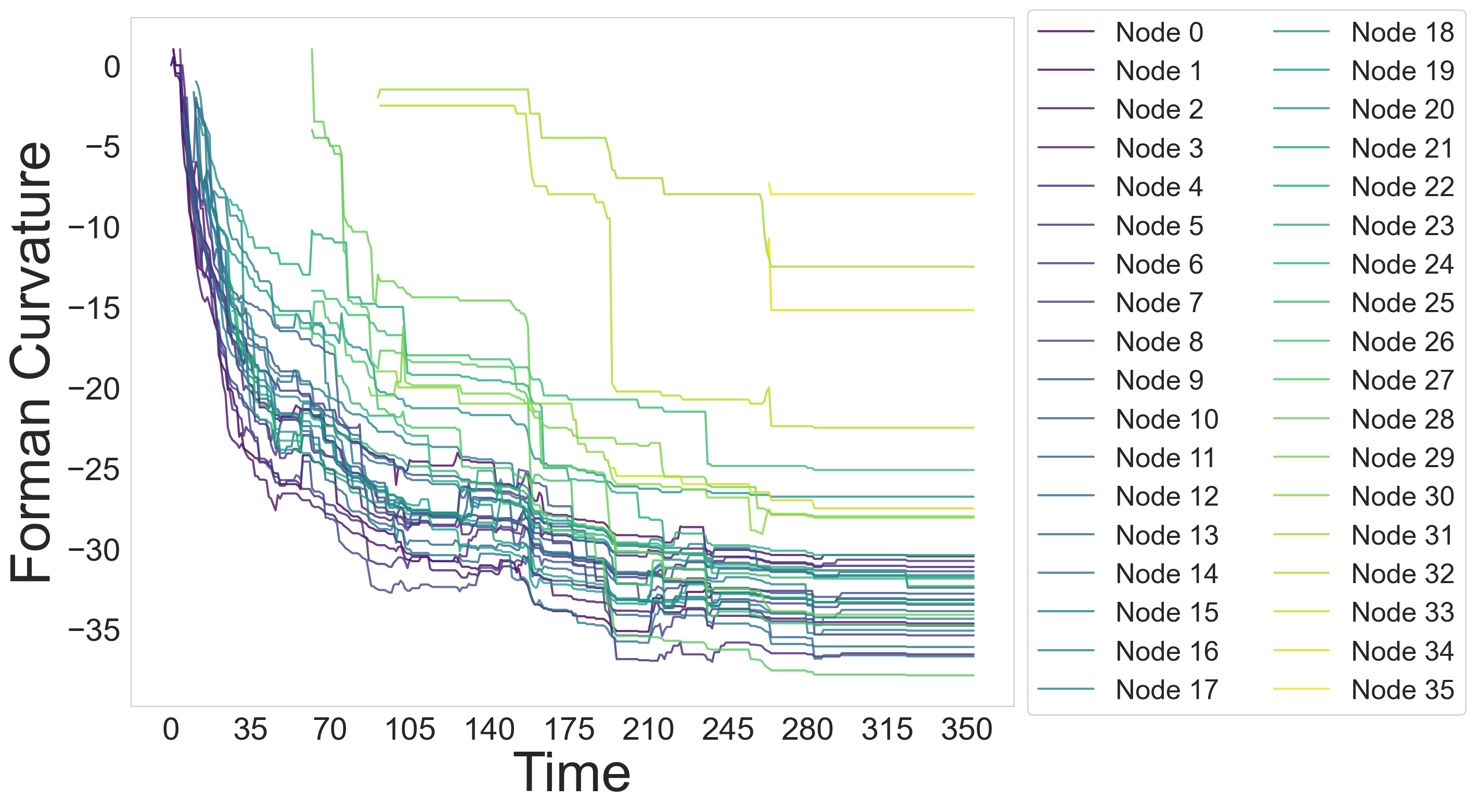}
            \caption{Evolution of Forman-Ricci node curvatures versus time for Fugue from Sonata No. 3}
            \label{fig:formanNodeCurv}
        \end{minipage}
    \end{figure}

\subsection{Distance in musical spaces}

\begin{figure}[H] 
        \centering
        \begin{minipage}[c]{0.49\textwidth} 
            \centering
        \includegraphics[width=\linewidth]{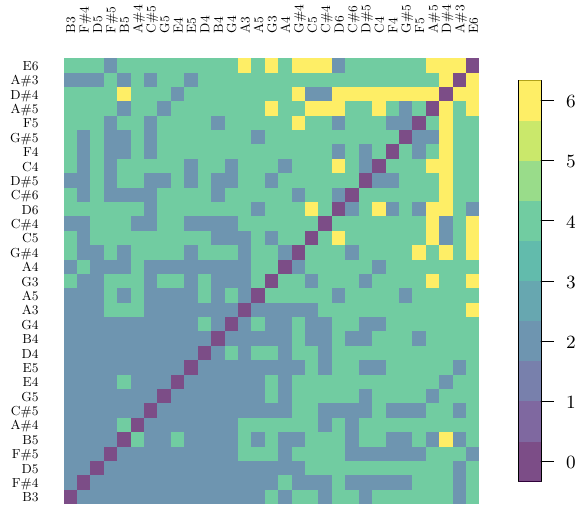}
            \caption{Visualization of the node to node pairwise distance using Connes's definition $d_{\mathcal{N},C}$ for partita 1 bourr\'ee.}
            \label{fig:connesBourree1}
        \end{minipage}
        \hfill
        \begin{minipage}[c]{0.49\textwidth} 
            \centering
        \includegraphics[width=1.35\linewidth]{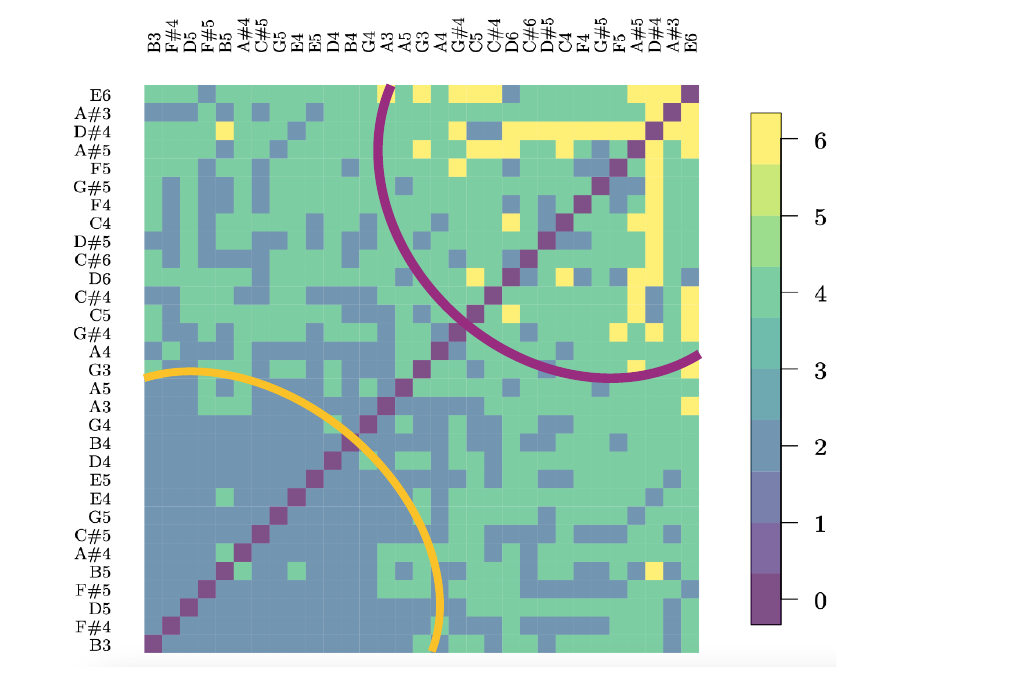}
            \caption{The orange and purple arcs in the figure are drawn to guide the eye to the identified regions around and away from the tonic, in which notes attract and repel each others respectively for partita 1 bourr\'ee.}
            \label{fig:Arcpartitabourree}
        \end{minipage}
    \end{figure}

\begin{figure}[H] 
        \centering
        \begin{minipage}[c]{0.49\textwidth} 
            \centering
        \includegraphics[width=\linewidth]{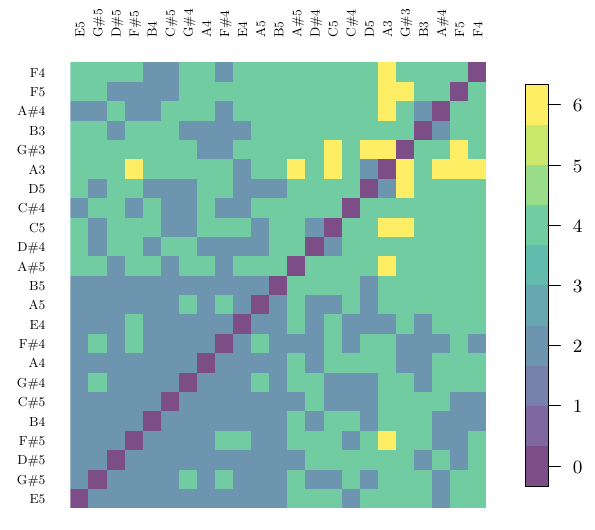}
            \caption{Visualization of the node to node pairwise distance using Connes's definition $d_{\mathcal{N},C}$ for partita 3 minuet I, II}
            \label{fig:connesMinuet}
        \end{minipage}
        \hfill
        \begin{minipage}[c]{0.49\textwidth} 
            \centering
        \includegraphics[width=1\linewidth]{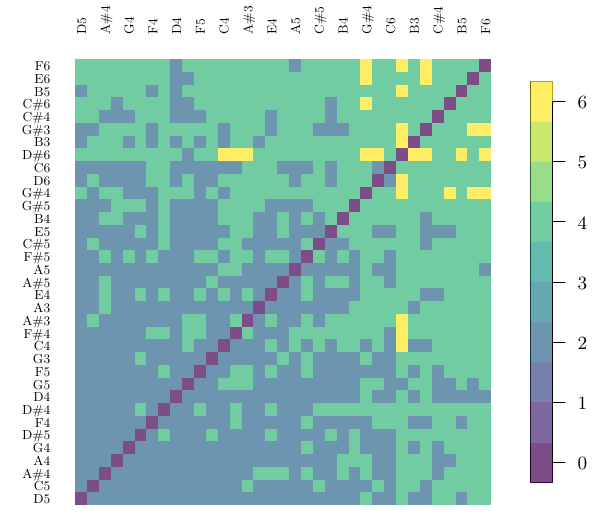}
            \caption{Visualization of the node to node pairwise distance using Connes's definition $d_{\mathcal{N},C}$ for sonata 1 fugue}
            \label{fig:connesFugue1}
        \end{minipage}
    \end{figure}

In order to recover distance in the composition space associated with every movement we solved Equation \ref{eqsimplex} using {\sf CVXR} package in {\sf R}. Our algorithm runs parallel threads to independently compute the distance for every pair and returns a matrix whose columns and rows are the individual notes and the values are their corresponding distances.  
The latter, Connes distances, are shown as heatmaps   (Figs. \ref{fig:connesBourree1}, \ref{fig:connesMinuet}, and \ref{fig:connesFugue1}). 
The musical notes on the heatmap are displayed in the sequence of their initial occurrence over time. In the lower left corner, we have the starting note of the musical movement, and as we go higher, the subsequent notes are those which progressively appear in the music as time advances. Generally, in music theory, the starting note of a musical piece is either the tonic or the dominant, which have corresponding roles in functional harmony, and clearly these are the starting notes of our musical movements.

First, we note that there is a lower part in the heatmap (Fig. \ref{fig:connesBourree1}), which is delineated by an orange arc in Fig. \ref{fig:Arcpartitabourree} where distances are relatively small compared to a maximum distance $d_{\mathcal{N},C} = 6$ recovered for this piece. Conversely, the upper part of the heatmap is delineated by the purple arc and corresponds to notes that are relatively repelled from each other with large pairwise distances. Those notes are those that appear later in the piece. We note that this pattern holds for all the movements considered. We show below the results for Minuet from Partita 3 in Fig. \ref{fig:connesMinuet}, and the Fugue from Sonata 1 in Fig. \ref{fig:connesFugue1}, while the rest are in the Appendix.

Two primary aspects of notes and pitches are present in the displayed heatmaps: pitch and note identity. Notes with varying degrees of functional importance occur in different registers (octaves) throughout the music. The heatmaps reveal how a piece's key signature and the violin's standard register shape its musical language.
For the Bourr\'ee composed in B minor (Fig. \ref{fig:connesBourree1}), the notes B, F\#, C\#, D, A, appear most frequently. In terms of register, Baroque violin writing tends to favor the lower-middle registers of the violin's range. In our heatmaps, our analytical approach centers around the lower-left quadrant, where notes falling within the lower-middle range and conforming to the key signature are sitting. These notes tend to show shorter distances and thus more interconnectedness. In contrast, the upper right quadrant reveals contextual outliers, where notes either correspond to high-register pitches or notes that deviate from the diatonic scale, being chromatically altered (C instead of C\# for Bourr\'ee, etc).

For these movements, we also show the pairwise distance defined as the geodesic over the simple graph using {\sf distances} of {\sf igraph}, which does not treat the movement as a simplicial complex but rather reduces it to its node-to-node interaction network. We note that the results reveal no functional roles of the notes using this definition. They are shown in the last section of Appendix.


\section{Conclusion}
What does this all mean? The recovered dimension, curvature, and distance allowed us to see in the mind's eye the composition space in which notes are positioned according to harmonic rules, which impose selection constraints on the composer. One can then naturally ask how the choice of successive notes/chords is then made? The answer we provide is an image of a composer initially starting at 0, the position of the dominant note or the tonic, and performing a random walk in a curved space of a given dimension, with scalar curvature favoring a direction over others, and proximity defining the next note/chord in that direction of movement. So, if a composer closes his/her eyes and start composing, they see something like Figure \ref{fig:connesBourree1}, or so we hope! 
We also stress that our proposed framework, though adopted on a music data set,  is not restricted to this particular application. It generalizes to higher-order networks, and thus we propose metric computations which are application agnostic and can be applied to any simplicial complex. They bring in geometric identifiers and measures to supplement the topological ones setting the ground for a Gauss-Bonnet-like relations. The proposed definition of distance provides a geometric embedding, which is a highly active field in network science, which started with embedding large networks in hyperbolic latent spaces offering a description of their observed topological metrics \cite{krioukov2010hyperbolic}, particularly their scale-free nature and clustering, which is not the case in our music networks (neither are they large, nor do they have scale-free degree distributions). Nonetheless, our definition applies to arbitrary network sizes and topologies. Hyperbolic embedding is now applied in neural networks \cite{ganea2018hyperbolic,liu2019hyperbolic}, optimal routing \cite{boguna2010sustaining,kleinberg2007geographic}, and we see our proposed definition being adopted to these contexts as it is grounded in non-commutative geometry, the home of these operators.  Moreover,  notions of curvature have recently been used as geometric measures in applications related to resilience, risk, and robustness, with evidence from the literature on how it can be indicative of transitions near criticality and serve as a predictive metric for collapse and fragility in financial and economic networks,  cancer, brain and epidemic and a variety of biological networks \cite{farooq2019network,sandhu2015graph,sandhu2016ricci,benincasa2020curvature,samal2021network,samal2021networkfragility,joharinad2023mathematical}. It was also shown to be relevant to community detection through the Ricci-flow, which is based on the observation that curvature dictates the proximity between nodes; If the edge connecting them is negative, then they belong to different communities as opposed to when the sign of the edge is positive \cite{sia2019ollivier,ni2019community}. 

In this work, we have showcased the use of dimension, curvature, and distance on a sample data set and provided a glimpse into the relevance of the geometrical measures that we proposed to the network science community, which we believe will be of use in a variety of contexts and a wide spectrum of applications.  
\section*{Acknowledgments} S.N,  D.M, and M.E are thankful to the discussions with Prof. Ali Chamseddine who guided us to the proper literature and to Dr. Ola Malaeb for the thorough reading of the manuscript and the feedback she provided. D.M acknowledges Research Assistantship (RA) support for this work from the Center for Advanced Mathematical Sciences (CAMS) at the American University of Beirut. 
\section*{Authors' contributions} S.N conceived the theory and implemented the algorithms for distance computation, D.M prepared the data and carried out the curvature computations, and M.E validated the results on geometrical simplicial complexes both analytically and numerically, and recovered the expected values in the continuum limit. 
\section*{Appendix}
\section*{Properties of the Laplace operator}\label{lapop}
The three operators $\mathbf{L_k^{\text{down}}}$, $\mathbf{L_k^{\text{up}}}$ and $\mathbf{L_k}$ can be diagonalized simultaneously, and their corresponding eigenvectors obey the following relations:

$$ \text{im}(\mathbf{L^{{\text{down}}}_{k}}) \subseteq \ker(\mathbf{L^{\text{{up}}}_{k}}), $$
$$ \text{im}(\mathbf{L^{{\text{up}}}_{k}}) \subseteq \ker(\mathbf{L^{{\text{down}}}_{k}}). $$
$$\ker(\mathbf{L_k}) = \ker(\mathbf{L^{\text{up}}_k})
\cap \ker(\mathbf{L^{\text{down}}_k}).
$$
The eigenvectors of \( \mathbf{L}^{k} \) are classified into the following categories:  
\begin{itemize}
    \item[-] An eigenvector is a simultaneous eigenvector of \( \mathbf{L}^{\text{down}}_{k} \) and $\mathbf{L}^{\text{up}}_{k}$ corresponding to a nonzero eigenvalue \( \lambda^{k} \) of $\mathbf{L}^{\text{down}}_{k}$ , and a zero eigenvalue of \( \mathbf{L}^{\text{up}}_{k} \).
    \item[-] An eigenvector is a simultaneous eigenvector of \( \mathbf{L}^{\text{down}}_{k} \) and $\mathbf{L}^{\text{up}}_{k}$ corresponding to a nonzero eigenvalue \( \lambda^{k} \) of $\mathbf{L}^{\text{up}}_{k}$ , and a zero eigenvalue of \( \mathbf{L}^{\text{down}}_{k} \).
    \item[-] An eigenvector is an eigenvector common to \( \mathbf{L}_{k} \), \( \mathbf{L}^{\text{down}}_{k} \), and \( \mathbf{L}^{\text{up}}_{k} \), corresponding to the eigenvalue \( \lambda^{k} = 0 \).
\end{itemize}
We provide a proof for the above in what follows. Consider an eigenvector $\boldsymbol{\xi}$ of \( \mathbf{L}^{\text{up}}_{k} \) corresponding to $\lambda \neq 0$. This is translated as: \begin{equation*}
\mathbf{L}^{\text{up}}_{k} \boldsymbol{\xi} = \mathbf{B}_{k+1}\mathbf{B}_{k+1}^T \boldsymbol{\xi} = \lambda \boldsymbol{\xi}
\quad \Rightarrow \quad
\boldsymbol{\xi} = \frac{1}{\lambda} \mathbf{B}_{k+1}\mathbf{B}_{k+1}^T \boldsymbol{\xi}
\quad  
\end{equation*}
Next, we apply $\mathbf{L}^{\text{down}}_{k}$, and obtain:
\begin{equation*}
\mathbf{L}^{\text{down}}_{k} \boldsymbol{\xi} = \mathbf{B}_{k}^T\mathbf{B}_{k} \boldsymbol{\xi}
= \frac{1}{\lambda}  \mathbf{B}_{k}^T\mathbf{B}_{k} \mathbf{B}_{k+1}\mathbf{B}_{k+1}^T \boldsymbol{\xi}
= \mathbf{0}
\end{equation*}

\section*{Distance: Constraints' Computations}\label{const}
In \cite{requardt2002dirac}, the following simple equality for operators on Hilbert spaces is invoked: 
$$
\|T\| = \|T^*\|
$$ in order to compute $|[D, f]\|$.  We thus use:
$$
\|[B, f]\| = \|[B, \bar{f}]\| = \|[B^{T}, f]\|,
$$
leading to a definition of our constraint in terms of the incidence matrix (or exterior derivative in the language of discrete differential geometry):
$$
\|[D, f]\| = \|[B, f]\|
$$
The first constraint given by the above equation is over oriented edges $[i,j]$, which is the same as that of the graph whose commutator is given by:  
$$
[B_0, f]\alpha(i,j) = B_0(f \alpha(i,j)) - f(B_0 \alpha(i,j)),
$$
$f \in \Omega^0$ is 0-form (scalar) and $\alpha \in \Omega^1$ is 1-form (co-vector or in simple terms a row vector as opposed to a column vector). Their multiplication should give $\Omega^1 \times \Omega_0 \rightarrow \Omega^1$, which means that $f$ has to be pulled up to the dimension of $\alpha$ through an averaging given by: $$
(f\alpha)([i,j]) = \frac{f_i + f_j}{2} \alpha([i,j]),
$$
which is used to find the commutator. 
which gives: 
$$
B_0(f\alpha(i,j))  = \frac{(f_i + f_j)}{2} (\alpha(j) - \alpha(i)),
$$
and
$$
f (B_0 \alpha(i,j))  = f (\alpha(j) - \alpha(i)) = f_j \alpha(j) - f_i \alpha(i).
$$
Then the commutator is: 
$$ [B_0, f]\alpha(i,j) =  \frac{(f_i + f_j)}{2} (\alpha(j) - \alpha(i)) - (f_j \alpha(j) - f_i \alpha(i)) $$

$$ = \frac{f_i}{2} \alpha(j) - \frac{f_i}{2} \alpha(i) + \frac{f_j}{2} \alpha(j) - \frac{f_j}{2} \alpha(i) - f_j \alpha(j) + f_i \alpha(i) $$

$$[B_0, f]\alpha(i,j) = \frac{(f_i - f_j)}{2}(\alpha(i) + \alpha(j))  $$

Similarly to what we did for an edge, for an oriented triangle, $[i,j,k]$, $f$ has to be pulled up to the dimension of $\omega \in \Omega^2$, the 2-form on the triangle, through an averaging given by: 
$$
(f\omega)([i,j]) = \frac{f_i + f_j + f_k}{3} \omega([i,j,k]),
$$
Finally, for an oriented tetrahedron $[i,j,k,l]$, $f$ has to be pulled to the dimension of $\beta \in \Omega^3$ through an averaging given by: 
$$
(f\beta)([i,j,k,l]) = \frac{f_i + f_j + f_k + f_l}{4} \beta([i,j,k,l]),
$$

\section*{Supplementary Results}
\subsection*{$\Lambda$ computation}
The values $\Lambda$, the smallest non-zero eigenvalue of the movement's super Lapalcian, evaluated at the end of the piece, are tabulated below. 
\begin{table}[H]
\centering
\caption{Lambda ($\Lambda$) values for the movements}
\label{tab:lambdas}
\begin{tabular}{l r}
\toprule
\textbf{Movement} & $\boldsymbol{\Lambda}$ \\
\midrule
Partita 1 Allemanda        & 0.697 \\
Partita 1 Bourrée          & 0.089 \\
Partita 1 Sarabande        & 0.253 \\
Partita 2 Sarabande        & 0.246 \\
Partita 3 Loure            & 0.312 \\
Partita 3 Minuet I, II     & 0.950 \\
Sonata 1 Adagio            & 0.327 \\
Sonata 1 Fugue             & 0.045 \\
Sonata 1 Siciliano         & 0.285 \\
Sonata 2 Grave             & 0.272 \\
Sonata 2 Fugue             & 0.050 \\
Sonata 2 Andante           & 0.226 \\
Sonata 3 Adagio            & 0.114 \\
Sonata 3 Fugue             & 0.167 \\
Sonata 3 Largo             & 0.765 \\
\bottomrule
\end{tabular}
\end{table}

\subsection*{Temporal evolution of average scalar curvature}
Here we show the evolution of a sample of average scalar curvatures compared with the Forman counterpart. 
\begin{figure}[H]
    \centering
    \begin{minipage}[t]{0.3\textwidth}
        \includegraphics[width=\linewidth]{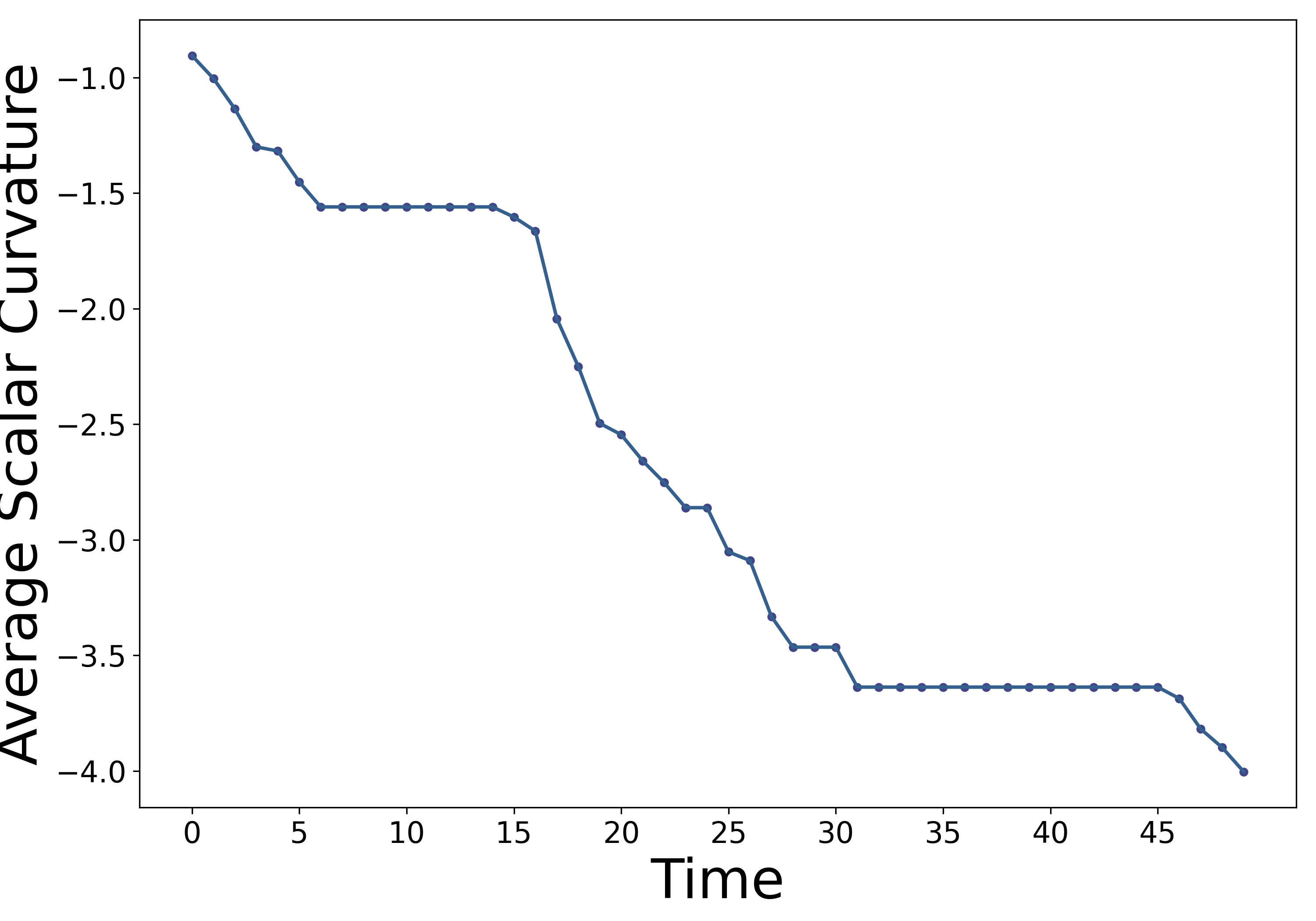}
        \caption{Average scalar curvature evolution using our proposed scalar curvature for dance movement Sarabande from Partita 2.}
        \label{lap-partita-2-sarabande}
    \end{minipage}
    \hspace{0.04 \textwidth}
    \begin{minipage}[t]{0.3\textwidth}
        \includegraphics[width=\linewidth]{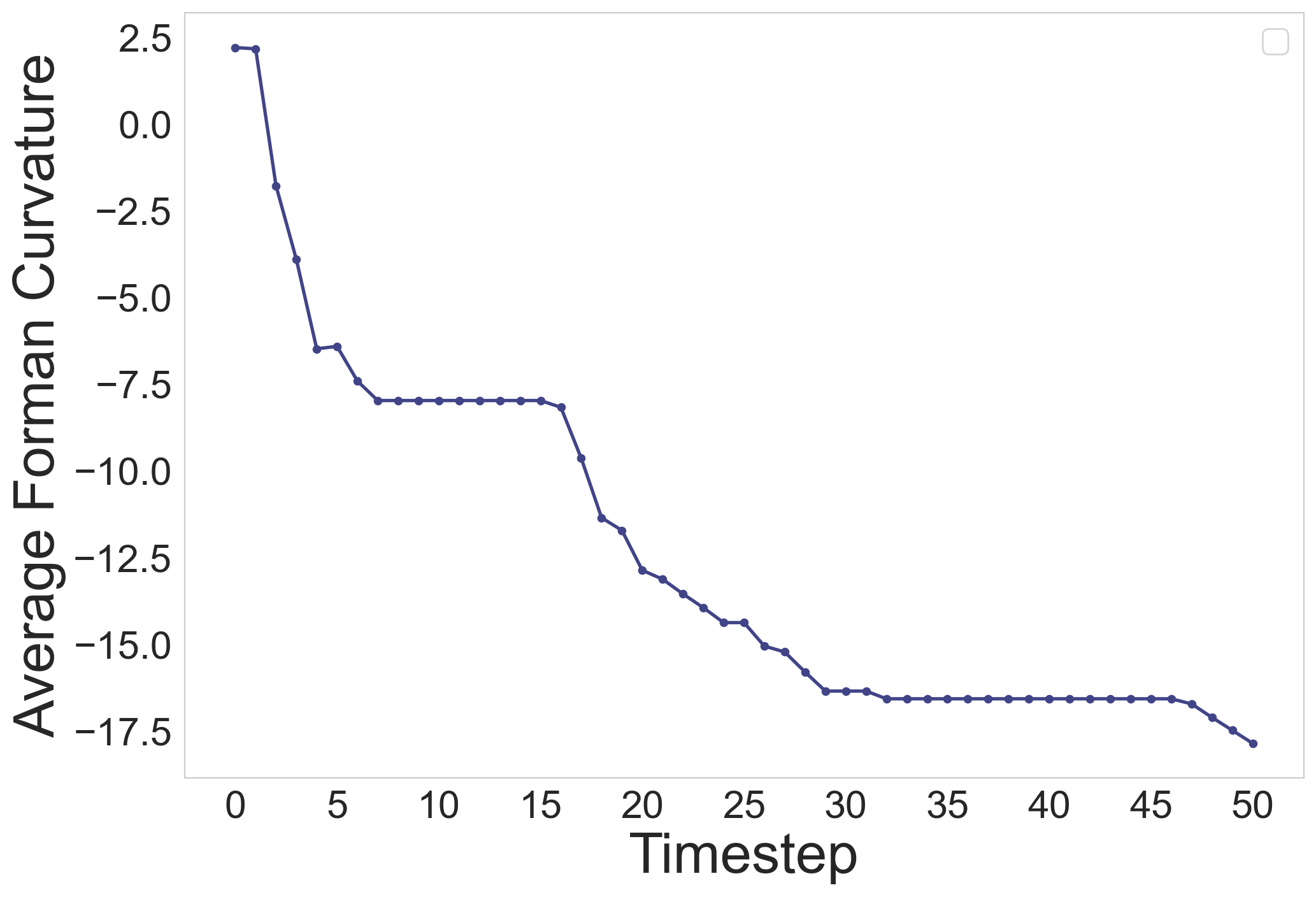}
        \caption{Average Forman-Ricci node curvature evolution for dance movement Sarabande from Partita 2.}
        \label{forman-partita-2-sarabande}
    \end{minipage}
\end{figure}

\subsection*{Connes distance in musical spaces}
We show below the results for rest of the movements for the recovered $d_{\mathcal{N},C}$. 
\begin{figure}[H]
    \centering
    \begin{minipage}[t]{0.3\textwidth}
        \includegraphics[width=\linewidth]{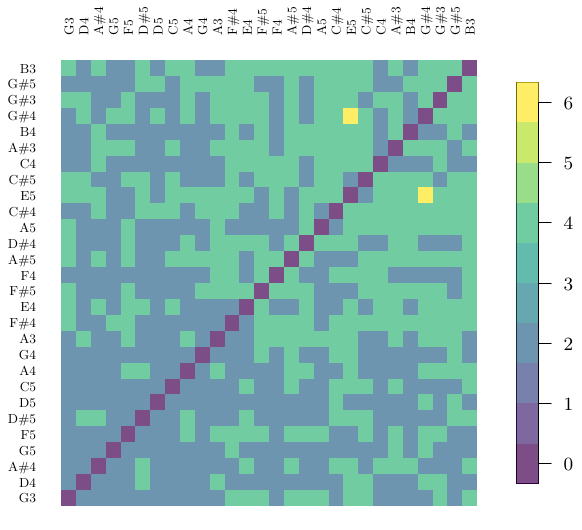}
        \caption{Sonata 1 Adagio}
        \label{distance-sonata-1-adagio}
    \end{minipage}
    \hfill
    \begin{minipage}[t]{0.3\textwidth}
        \includegraphics[width=\linewidth]{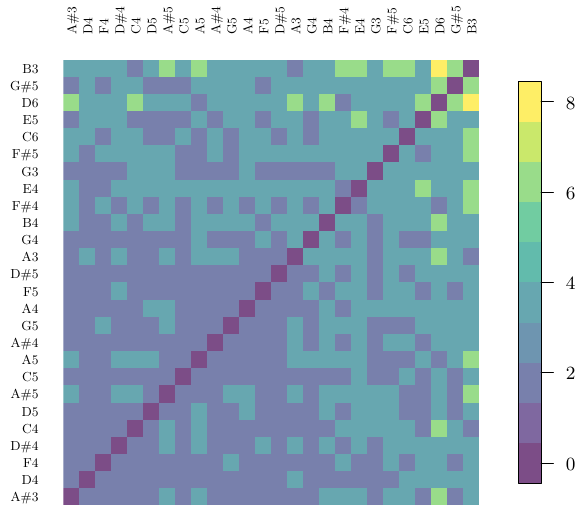}
        \caption{Sonata 1 Siciliano}
        \label{distance-sonata-1-siciliano}
    \end{minipage}
    \hfill
    \begin{minipage}[t]{0.3\textwidth}
        \includegraphics[width=\linewidth]{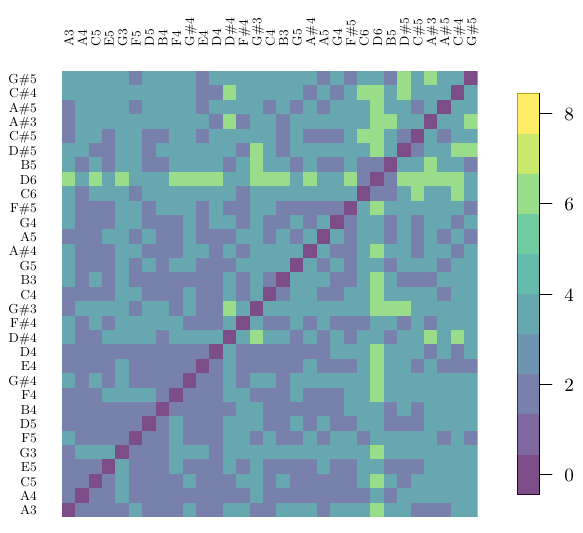}
        \caption{Sonata 2 Grave}
        \label{distance-sonata-2-grave}
    \end{minipage}
\end{figure}

\begin{figure}[H]
    \centering
    \begin{minipage}[t]{0.3\textwidth}
        \includegraphics[width=\linewidth]{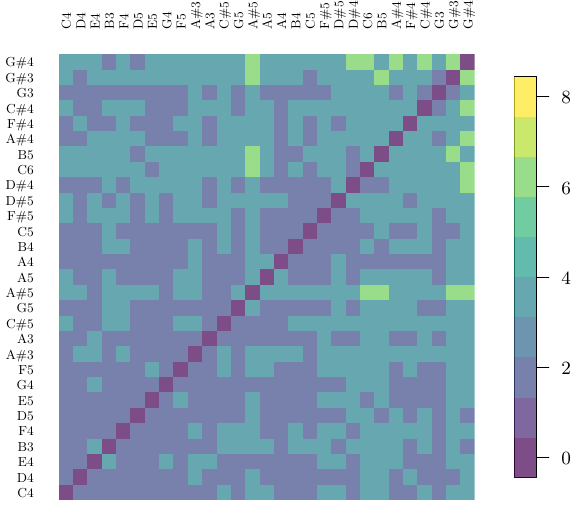}
        \caption{Sonata 3 Adagio}
        \label{distance-sonata-3-adagio}
    \end{minipage}
    \hfill
    \begin{minipage}[t]{0.3\textwidth}
        \includegraphics[width=\linewidth]{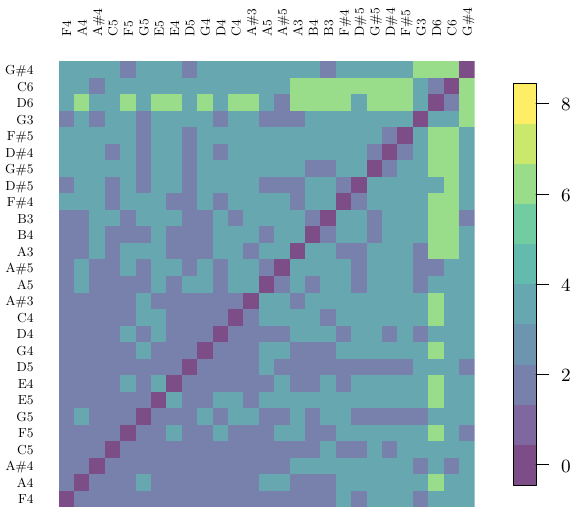}
        \caption{Sonata 3 Largo}
        \label{distance-sonata-3-largo}
    \end{minipage}
    \hfill
    \begin{minipage}[t]{0.3\textwidth}
        \includegraphics[width=\linewidth]{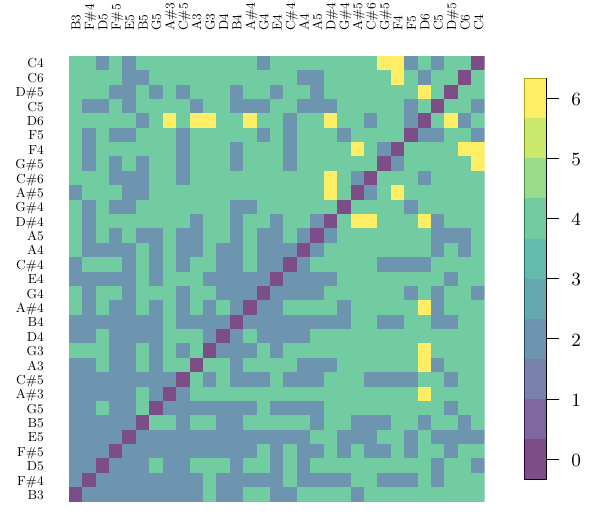}
        \caption{Partita 1 Allemanda}
        \label{distance-partita-1-allemanda}
    \end{minipage}
\end{figure}

\begin{figure}[H]
    \centering
    \begin{minipage}[t]{0.3\textwidth}
        \includegraphics[width=\linewidth]{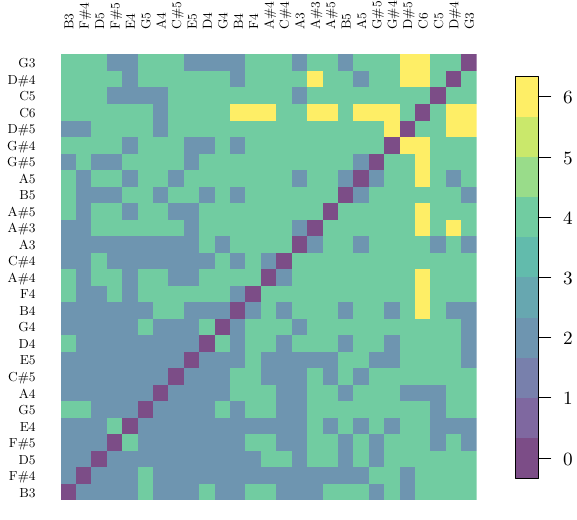}
        \caption{Partita 1 Sarabande}
        \label{distance-partita-1-sarabande}
    \end{minipage}
    \hfill
    \begin{minipage}[t]{0.3\textwidth}
        \includegraphics[width=\linewidth]{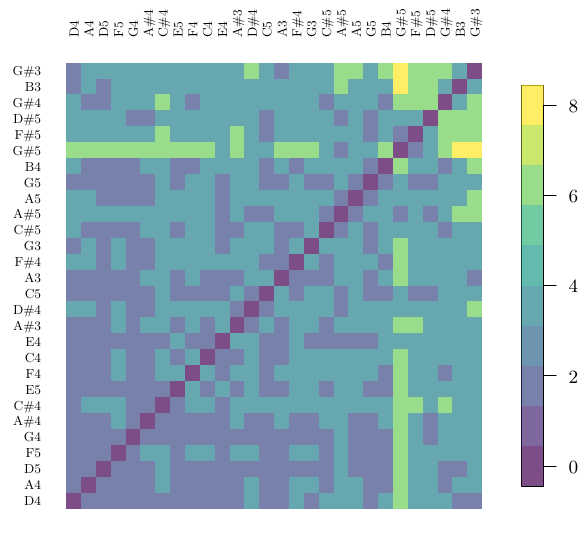}
        \caption{Partita 2 Sarabande}
        \label{distance-partita-2-sarabande}
    \end{minipage}
    \hfill
    \begin{minipage}[t]{0.3\textwidth}
        \includegraphics[width=\linewidth]{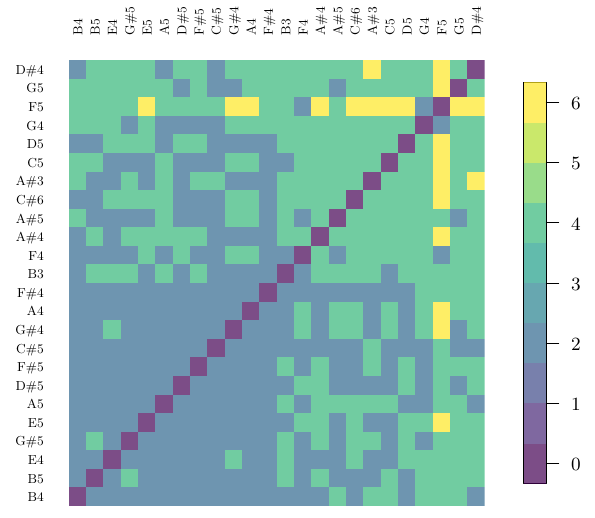}
        \caption{Partita 3 Loure}
        \label{distance-partita-3-loure}
    \end{minipage}
\end{figure}

\begin{figure}[H]
    \centering
    \begin{minipage}[t]{0.3\textwidth}
        \includegraphics[width=\linewidth]{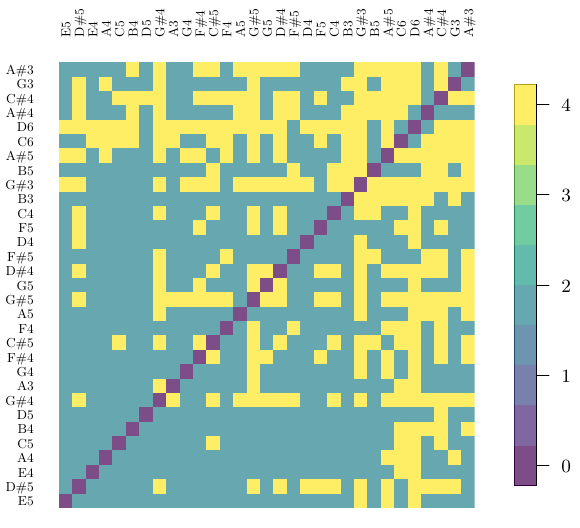}
        \caption{Sonata 2 Fugue}
        \label{distance-sonata-2-fugue}
    \end{minipage}
    \hspace{0.04 \textwidth}
    \begin{minipage}[t]{0.3\textwidth}
        \includegraphics[width=\linewidth]{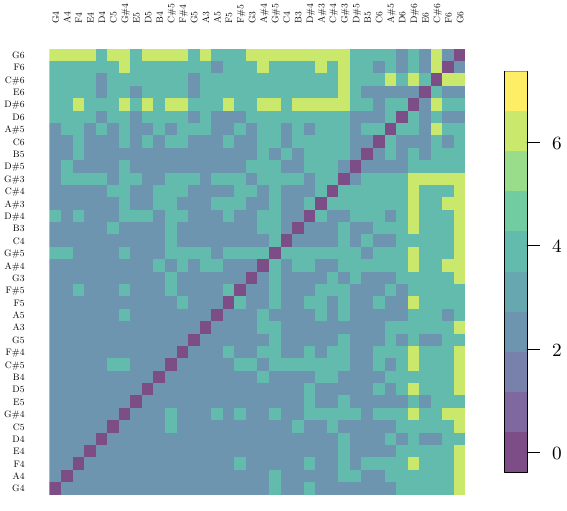}
        \caption{Sonata 3 Fugue}
        \label{distance-sonata-3-fugue}
    \end{minipage}
\end{figure}

\subsection*{Pairwise distance}

   We show the pairwise distances between notes defined as the geodesic over the simple graph using {\sf distances} of {\sf igraph}, which does not use the Connes distance. It's a Djikistra algorithm over the graph. 
\begin{figure}[H]
    \centering
    \begin{minipage}[t]{0.3\textwidth}
        \includegraphics[width=\linewidth]{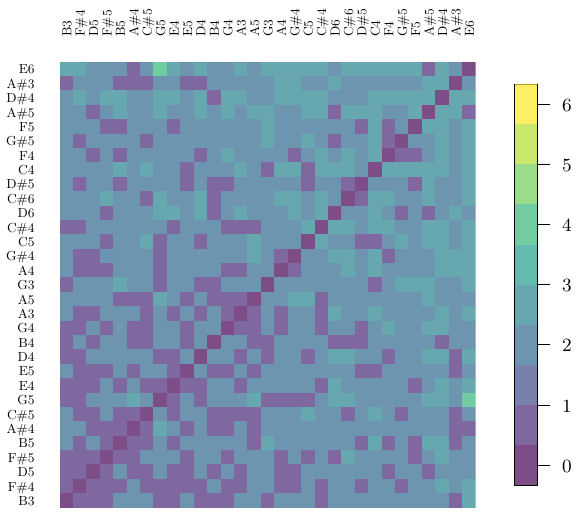}
        \caption{Partita 1 Bourr\'ee}
        \label{pairwise-partita-1-bourree}
    \end{minipage}
    \hfill
    \begin{minipage}[t]{0.3\textwidth}
        \includegraphics[width=\linewidth]{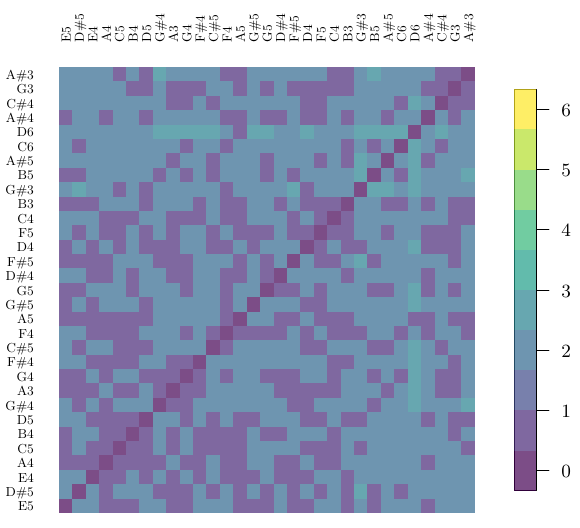}
        \caption{Sonata 2 Fugue}
        \label{pairwise-sonata-2-fugue}
    \end{minipage}
    \hfill
    \begin{minipage}[t]{0.3\textwidth}
        \includegraphics[width=\linewidth]{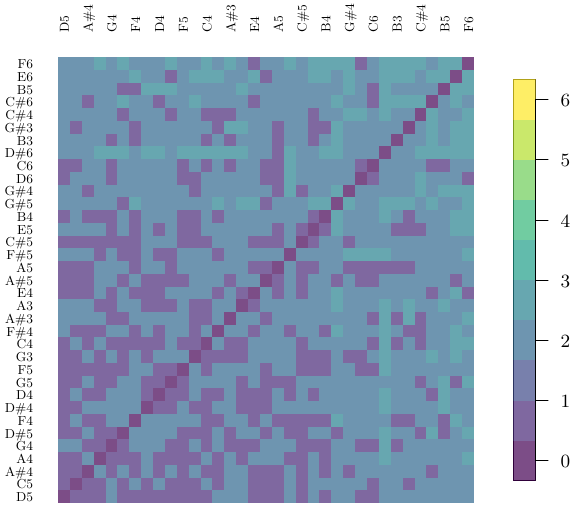}
        \caption{Sonata 1 Fugue}
        \label{pairwise-sonata-1-fugue}
    \end{minipage}
\end{figure}

\bibliographystyle{plain}
\bibliography{ref}

\begin{thebibliography}{10}

\bibitem{aref2020modeling}
Samin Aref, Andrew~J Mason, and Mark~C Wilson.
\newblock A modeling and computational study of the frustration index in signed
  networks.
\newblock {\em Networks}, 75(1):95--110, 2020.

\bibitem{battiston2021physics}
Federico Battiston, Enrico Amico, Alain Barrat, Ginestra Bianconi, Guilherme
  Ferraz~de Arruda, Benedetta Franceschiello, Iacopo Iacopini, Sonia K{\'e}fi,
  Vito Latora, Yamir Moreno, et~al.
\newblock The physics of higher-order interactions in complex systems.
\newblock {\em Nature physics}, 17(10):1093--1098, 2021.

\bibitem{battiston2022higher}
Federico Battiston and Giovanni Petri.
\newblock {\em Higher-order systems}.
\newblock Springer, 2022.

\bibitem{bauer2011ollivier}
Frank Bauer, J{\"u}rgen Jost, and Shiping Liu.
\newblock Ollivier-ricci curvature and the spectrum of the normalized graph
  laplace operator.
\newblock {\em arXiv preprint arXiv:1105.3803}, 2011.

\bibitem{benincasa2020curvature}
D.~M.~T. Benincasa and F.~Dowker.
\newblock Using curvature to infer covid-19 epidemic network fragility and
  systemic risk.
\newblock {\em Journal of Statistical Mechanics: Theory and Experiment},
  2021:053501, 2020.

\bibitem{benson2016higher}
Austin~R Benson, David~F Gleich, and Jure Leskovec.
\newblock Higher-order organization of complex networks.
\newblock {\em Science}, 353(6295):163--166, 2016.

\bibitem{besnard2021estimating}
Fabien Besnard.
\newblock Estimating noncommutative distances on graphs.
\newblock {\em arXiv preprint arXiv:2105.09056}, 2021.

\bibitem{bianconi2021higher}
Ginestra Bianconi.
\newblock {\em Higher-order networks}.
\newblock Cambridge University Press, 2021.

\bibitem{bianconi2021topological}
Ginestra Bianconi.
\newblock The topological dirac equation of networks and simplicial complexes.
\newblock {\em Journal of Physics: Complexity}, 2(3):035022, 2021.

\bibitem{boguna2010sustaining}
Mari{\'a}n Bogun{\'a}, Fragkiskos Papadopoulos, and Dmitri Krioukov.
\newblock Sustaining the internet with hyperbolic mapping.
\newblock {\em Nature communications}, 1(1):62, 2010.

\bibitem{burioni1996universal}
Raffaella Burioni and Davide Cassi.
\newblock Universal properties of spectral dimension.
\newblock {\em Physical review letters}, 76(7):1091, 1996.

\bibitem{buser2010geometry}
Peter Buser.
\newblock {\em Geometry and spectra of compact Riemann surfaces}.
\newblock Springer Science \& Business Media, 2010.

\bibitem{calmon2022higher}
Lucille Calmon, Michael~T Schaub, and Ginestra Bianconi.
\newblock Higher-order signal processing with the dirac operator.
\newblock In {\em 2022 56th Asilomar Conference on Signals, Systems, and
  Computers}, pages 925--929. IEEE, 2022.

\bibitem{calmon2023dirac}
Lucille Calmon, Michael~T Schaub, and Ginestra Bianconi.
\newblock Dirac signal processing of higher-order topological signals.
\newblock {\em New Journal of Physics}, 25(9):093013, 2023.

\bibitem{chakraborty2025shortest}
Sukrit Chakraborty, Prasanta Choudhury, and Arindam Mukherjee.
\newblock Shortest paths in a weighted simplicial complex.
\newblock {\em arXiv preprint arXiv:2506.12921}, 2025.

\bibitem{chamseddine1997spectral}
Ali~H Chamseddine and Alain Connes.
\newblock The spectral action principle.
\newblock {\em Communications in Mathematical Physics}, 186(3):731--750, 1997.

\bibitem{chavel1984eigenvalues}
Isaac Chavel.
\newblock {\em Eigenvalues in Riemannian geometry}, volume 115.
\newblock Academic press, 1984.

\bibitem{connes2013spectral}
Alain Connes.
\newblock On the spectral characterization of manifolds.
\newblock {\em Journal of Noncommutative Geometry}, 7(1):1--82, 2013.

\bibitem{connes2010noncommutative}
Alain Connes, Joachim Cuntz, and Marc~A Rieffel.
\newblock Noncommutative geometry.
\newblock {\em Oberwolfach Reports}, 6(3):2271--2334, 2010.

\bibitem{connes1995local}
Alain Connes and Henri Moscovici.
\newblock The local index formula in noncommutative geometry.
\newblock {\em Geometric \& Functional Analysis GAFA}, 5(2):174--243, 1995.

\bibitem{dimakis1998connes}
Aristophanes Dimakis and Folkert Muller-Hoissen.
\newblock Connes' distance function on one-dimensional lattices.
\newblock {\em International journal of theoretical physics}, 37(3):907--913,
  1998.

\bibitem{durhuus2007spectral}
Bergfinnur Durhuus, Thordur Jonsson, and John~F Wheater.
\newblock The spectral dimension of generic trees.
\newblock {\em Journal of Statistical Physics}, 128(5):1237--1260, 2007.

\bibitem{farooq2019network}
Hamza Farooq, Yongxin Chen, Tryphon~T Georgiou, Allen Tannenbaum, and
  Christophe Lenglet.
\newblock Network curvature as a hallmark of brain structural connectivity.
\newblock {\em Nature communications}, 10(1):4937, 2019.

\bibitem{forman2003bochner}
Forman.
\newblock Bochner's method for cell complexes and combinatorial ricci
  curvature.
\newblock {\em Discrete \& Computational Geometry}, 29(3):323--374, 2003.

\bibitem{ganea2018hyperbolic}
Octavian Ganea, Gary B{\'e}cigneul, and Thomas Hofmann.
\newblock Hyperbolic neural networks.
\newblock {\em Advances in neural information processing systems}, 31, 2018.

\bibitem{hwang2010spectral}
S~Hwang, C-K Yun, D-S Lee, B~Kahng, and D~Kim.
\newblock Spectral dimensions of hierarchical scale-free networks with weighted
  shortcuts.
\newblock {\em Physical Review E—Statistical, Nonlinear, and Soft Matter
  Physics}, 82(5):056110, 2010.

\bibitem{joharinad2023mathematical}
P.~Joharinad and J.~Jost.
\newblock {\em Mathematical Principles of Topological and Geometric Data
  Analysis}.
\newblock Mathematics of Data. Springer, 2023.

\bibitem{khalkhali2013basic}
Masoud Khalkhali.
\newblock {\em Basic noncommutative geometry}.
\newblock European mathematical society Z{\"u}rich, Switzerland, 2013.

\bibitem{khalkhali2008invitation}
Masoud Khalkhali and Matilde Marcolli.
\newblock {\em An invitation to noncommutative geometry}.
\newblock World Scientific, 2008.

\bibitem{kleinberg2007geographic}
Robert Kleinberg.
\newblock Geographic routing using hyperbolic space.
\newblock In {\em IEEE INFOCOM 2007-26th IEEE International Conference on
  Computer Communications}, pages 1902--1909. IEEE, 2007.

\bibitem{krioukov2010hyperbolic}
Dmitri Krioukov, Fragkiskos Papadopoulos, Maksim Kitsak, Amin Vahdat, and
  Mari{\'a}n Bogun{\'a}.
\newblock Hyperbolic geometry of complex networks.
\newblock {\em Physical Review E—Statistical, Nonlinear, and Soft Matter
  Physics}, 82(3):036106, 2010.

\bibitem{lambiotte2019networks}
Renaud Lambiotte, Martin Rosvall, and Ingo Scholtes.
\newblock From networks to optimal higher-order models of complex systems.
\newblock {\em Nature physics}, 15(4):313--320, 2019.

\bibitem{leal2021forman}
Wilmer Leal, Guillermo Restrepo, Peter~F Stadler, and J{\"u}rgen Jost.
\newblock Forman--ricci curvature for hypergraphs.
\newblock {\em Advances in Complex Systems}, 24(01):2150003, 2021.

\bibitem{li2023hyperbolic}
Jinming Li, Gongjun Xu, and Ji~Zhu.
\newblock Hyperbolic network latent space model with learnable curvature.
\newblock {\em arXiv preprint arXiv:2312.05319}, 2023.

\bibitem{lin2011ricci}
Yong Lin, Linyuan Lu, and Shing-Tung Yau.
\newblock Ricci curvature of graphs.
\newblock {\em Tohoku Mathematical Journal, Second Series}, 63(4):605--627,
  2011.

\bibitem{liu2019hyperbolic}
Qi~Liu, Maximilian Nickel, and Douwe Kiela.
\newblock Hyperbolic graph neural networks.
\newblock {\em Advances in neural information processing systems}, 32, 2019.

\bibitem{mayfield2017higher}
Margaret~M Mayfield and Daniel~B Stouffer.
\newblock Higher-order interactions capture unexplained complexity in diverse
  communities.
\newblock {\em Nature ecology \& evolution}, 1(3):0062, 2017.

\bibitem{millan2025topology}
Ana~P Mill{\'a}n, Hanlin Sun, Lorenzo Giambagli, Riccardo Muolo, Timoteo
  Carletti, Joaqu{\'\i}n~J Torres, Filippo Radicchi, J{\"u}rgen Kurths, and
  Ginestra Bianconi.
\newblock Topology shapes dynamics of higher-order networks.
\newblock {\em Nature Physics}, pages 1--9, 2025.

\bibitem{millan2022geometry}
Ana~Paula Mill{\'a}n, Juan~G Restrepo, Joaqu{\'\i}n~J Torres, and Ginestra
  Bianconi.
\newblock Geometry, topology and simplicial synchronization.
\newblock In {\em Higher-Order Systems}, pages 269--299. Springer, 2022.

\bibitem{minakshisundaram1949some}
Subbaramiah Minakshisundaram and {\AA}ke Pleijel.
\newblock Some properties of the eigenfunctions of the laplace-operator on
  riemannian manifolds.
\newblock {\em Canadian Journal of Mathematics}, 1(3):242--256, 1949.

\bibitem{mrad2025higher}
Dima Mrad and Sara Najem.
\newblock Higher-order network representation of js bach's solo violin sonatas
  and partitas: Topological and geometrical explorations.
\newblock {\em arXiv preprint arXiv:2506.08540}, 2025.

\bibitem{ni2019community}
Chien-Chun Ni, Yu-Yao Lin, Feng Luo, and Jie Gao.
\newblock Community detection on networks with ricci flow.
\newblock {\em Scientific reports}, 9(1):9984, 2019.

\bibitem{nortier2025higher}
Bern{\'e}~L Nortier, Simon Dobson, and Federico Battiston.
\newblock Higher-order shortest paths in hypergraphs.
\newblock {\em arXiv preprint arXiv:2502.03020}, 2025.

\bibitem{ollivier2009ricci}
Yann Ollivier.
\newblock Ricci curvature of markov chains on metric spaces.
\newblock {\em Journal of Functional Analysis}, 256(3):810--864, 2009.

\bibitem{requardt1997new}
Manfred Requardt.
\newblock A new approach to functional analysis on graphs, the connes-spectral
  triple and its distance function.
\newblock {\em arXiv preprint hep-th/9708010}, 1997.

\bibitem{requardt2000graph}
Manfred Requardt.
\newblock Graph-laplacians and dirac operators on (infinite) graphs and the
  calculation of the connes-distance-functional.
\newblock {\em arXiv preprint math-ph/0001026}, 2000.

\bibitem{requardt2002dirac}
Manfred Requardt.
\newblock Dirac operators and the calculation of the connes metric on arbitrary
  (infinite) graphs.
\newblock {\em Journal of Physics A: Mathematical and General}, 35(3):759,
  2002.

\bibitem{samal2021networkfragility}
A.~Samal, S.~Kumar, Y.~Yadav, and A.~Chakraborti.
\newblock Network-centric indicators for fragility in global financial indices.
\newblock {\em Frontiers in Physics}, 8:624373, 2021.

\bibitem{samal2021network}
A.~Samal, H.~K. Pharasi, S.~J. Ramaia, H.~Kannan, E.~Saucan, and J.~Jost.
\newblock Network geometry and market instability.
\newblock {\em Royal Society Open Science}, 8(2):201734, 2021.

\bibitem{sandhu2016ricci}
R.~Sandhu, T.~T. Georgiou, and A.~R. Tannenbaum.
\newblock Ricci curvature: An economic indicator for market fragility and
  systemic risk.
\newblock {\em Science Advances}, 2(5):e1501495, 2016.

\bibitem{sandhu2015graph}
Romeil Sandhu, Tryphon Georgiou, Ed~Reznik, Liangjia Zhu, Ivan Kolesov, Yasin
  Senbabaoglu, and Allen Tannenbaum.
\newblock Graph curvature for differentiating cancer networks.
\newblock {\em Scientific reports}, 5(1):12323, 2015.

\bibitem{scholtes2016higher}
Ingo Scholtes, Nicolas Wider, and Antonios Garas.
\newblock Higher-order aggregate networks in the analysis of temporal networks:
  path structures and centralities.
\newblock {\em The European Physical Journal B}, 89(3):61, 2016.

\bibitem{sia2019ollivier}
Jayson Sia, Edmond Jonckheere, and Paul Bogdan.
\newblock Ollivier-ricci curvature-based method to community detection in
  complex networks.
\newblock {\em Scientific reports}, 9(1):9800, 2019.

\bibitem{sreejith2016forman}
Remanan~Pushpa Sreejith, Karthikeyan Mohanraj, J{\"u}rgen Jost, Emil Saucan,
  and Areejit Samal.
\newblock Forman curvature for complex networks.
\newblock {\em Journal of Statistical Mechanics: Theory and Experiment},
  2016(6):063206, 2016.

\bibitem{torres2020simplicial}
Joaqu{\'\i}n~J Torres and Ginestra Bianconi.
\newblock Simplicial complexes: higher-order spectral dimension and dynamics.
\newblock {\em Journal of Physics: Complexity}, 1(1):015002, 2020.

\bibitem{vasilyeva2023distances}
Ekaterina Vasilyeva, Miguel Romance, Ivan Samoylenko, Kirill Kovalenko, Daniil
  Musatov, Andrey~Mihailovich Raigorodskii, and Stefano Boccaletti.
\newblock Distances in higher-order networks and the metric structure of
  hypergraphs.
\newblock {\em Entropy}, 25(6):923, 2023.

\bibitem{weber2017characterizing}
Melanie Weber, Emil Saucan, and J{\"u}rgen Jost.
\newblock Characterizing complex networks with forman-ricci curvature and
  associated geometric flows.
\newblock {\em Journal of Complex Networks}, 5(4):527--550, 2017.

\end{thebibliography}

\end{document}